\documentclass[12pt]{iopart}
\usepackage{graphicx}

\begin{document}

\topical[Magnetism in Fe-based superconductors]{Magnetism in Fe-based superconductors}

\author{M.D. Lumsden and A.D. Christianson}

\address{Neutron Scattering Science Division, Oak Ridge National Laboratory,
Oak Ridge, TN, 37831, USA.}
\eads{lumsdenmd@ornl.gov, christiansonad@ornl.gov}

\begin{abstract}
In this review, we present a summary of experimental studies of magnetism in Fe-based superconductors.  The doping dependent phase diagram shows strong similarities to the generic phase diagram of the cuprates.  Parent compounds exhibit magnetic order together with a structural phase transition both of which are progressively suppressed with doping allowing superconductivity to emerge.  The stripe-like spin arrangement of Fe moments in the magnetically ordered state shows the identical in-plane structure for the RFeAsO (R=rare earth) and AFe$_2$As$_2$ (A=Sr, Ca, Ba, Eu and K) parent compounds, notably different than the spin configuration of the cuprates.  Interestingly, Fe$_{1+y}$Te orders with a different spin order despite very similar Fermi surface topology.  Studies of the spin dynamics in the parent compounds shows that the interactions are best characterized as anisotropic three-dimensional (3D) interactions.  Despite the room temperature tetragonal structure, analysis of the low temperature spin waves under the assumption of a Heisenberg Hamiltonian indicates strong in-plane anisotropy with a significant next-near neighbor interaction.  In the superconducting state, a resonance, localized in both wavevector and energy is observed in the spin excitation spectrum as in the cuprates. This resonance is observed at a wavevector compatible with a Fermi surface nesting instability independent of the magnetic ordering of the relevant parent compound. The resonance energy (E$_r$) scales with superconducting transition temperature (T$_C$) as E$_r$ $\sim$4.9 k$_B$T$_C$ consistent with the canonical value of $\sim$5 k$_B$T$_C$ observed in the cuprates.  Moreover, the relationship between the resonance energy and the superconducting gap, $\Delta$, is similar to that observed in many unconventional superconductors (E$_r$/2$\Delta$ $\sim$ 0.64).

\end{abstract}



\tableofcontents 

\maketitle

\section{Introduction}
The discovery of superconductivity in F-doped LaFeAsO with T$_C$ of 26 K \cite{Kamihara2008c} created a flurry of excitement in the condensed matter physics community.  Substitutional replacement of the rare earth ion led to a rapid increase in the superconducting transition temperature.  Denoting the chemical formula of these so-called 1111-materials by RFeAsO, F-doping resulted in the following optimal transition temperatures: 52 K for R=Nd\cite{Ren2008e}, 52 K for R=Pr\cite{Ren2008d}, 55 K for R=Sm\cite{Ren2008c}, 41 K For R=Ce \cite{Chen2008d}, 36-50 K for R=Gd \cite{Cheng2008,Kadowaki2009,Khlybov2009}, 46 K for R=Tb \cite{Bos2008}, and 45 K for R=Dy \cite{Bos2008}. To date, the highest transition temperature for the Fe-based superconductors is 56 K observed in a sample of Gd$_{1-x}$Th$_x$FeAsO \cite{Wang2008}. These superconducting transition temperatures make the Fe-based materials second only to the cuprates and they, therefore, represent the second family of high-T$_C$ superconductors.  It was later shown that F doping was not necessary and similar transition temperatures could be obtained for the case of oxygen deficient RFeAsO$_{1-y}$ : 28 K for R=La \cite{Ren2008b,Miyazawa2009a}, 42 K for Ce \cite{Ren2008b}, 53 K for Nd \cite{Ren2008b,Miyazawa2009a,Kito2008}, 48 K for Pr \cite{Ren2008b,Miyazawa2009a}, 55 K for Sm \cite{Ren2008b}, 53 K for Gd \cite{Yang2008c,Miyazawa2009a}, 52 K for Tb and Dy \cite{Miyazawa2009a}. Both the F doped and oxygen deficient samples show the same trend for T$_C$ as a function of rare earth ion.


\begin{figure}
\includegraphics[width=\columnwidth]{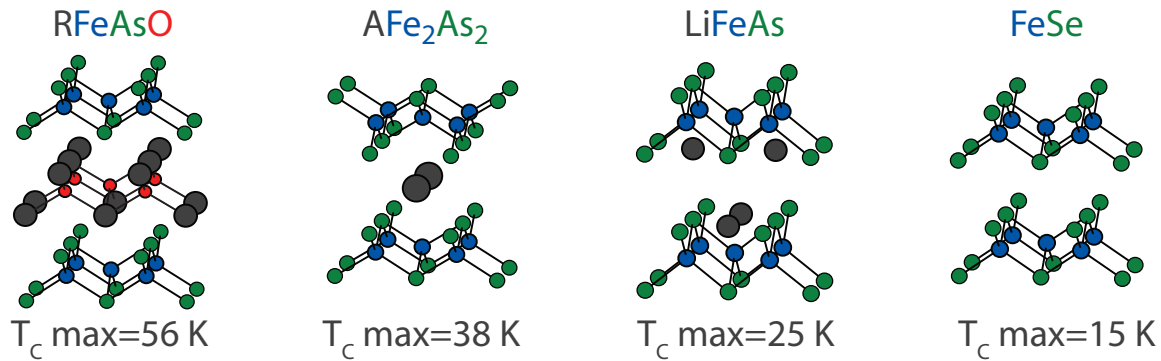}
\caption{Crystal structure of the 1111, 122, 111, and 11 materials.  In all cases, the FeAs (or FeSe) plane is the same with the principle difference being the spacer between layers.  Below each figure is the maximum T$_C$ observed at ambient pressure in each of the families.  From the figure, it is clear that the larger the separation between layers, the higher the transition temperature. (Figure courtesy M.A. McGuire) }
\label{structure}
\end{figure}

At room temperature, the RFeAsO materials crystallize in the $P4/nmm$ tetragonal space group resulting in a layered structure with FeAs and RO layers \cite{Cruz2008} (see Fig. \ref{structure}).  This layered structure is reminiscent of the cuprates with FeAs planes taking the place of the CuO layers.  The square planar arrangement of (likely) magnetic Fe is similar to the cuprates and leads one to naturally speculate that magnetism may play as essential role in the superconducting pairing.  Shortly after the initial activity on the 1111 materials, superconductivity was also discovered in related materials possessing identical FeAs layers with differing spacers.  The discovery of superconductivity with T$_C$ of 38 K in Ba$_{1-x}$K$_x$Fe$_2$As$_2$ \cite{Rotter2008a} was of particular interest as it was quickly realized that large single crystals of these 122 materials could be grown (unlike the 1111 family).  Structurally, at room temperature, the 122 materials exhibit the ThCr$_2$Si$_2$ crystal structure (space group $I4/mmm$).  As is clearly shown in Fig. \ref{structure}, the FeAs layers are very similar to the 1111 materials although neighboring layers along the $c$-axis have an inverted arsenic coordination.  For both the 1111\cite{Sefat2008a} and 122\cite{Sefat2008b} materials, it was quickly realized that replacement of Fe with Co would also result in superconductivity, albeit with a reduced T$_C$ when compared with doping between FeAs planes. Such behavior is in contrast to that observed in the cuprates where disorder in the copper oxide plane was found to destroy superconductivity.  Superconductivity was also discovered in the 122 materials upon electron doping on the Fe site with Ni \cite{Li2009d}, Rh \cite{Ni2009, Han2009}, Ir \cite{Han2009} and Pd \cite{Ni2009, Han2009} or by isoelectronic replacement of Fe with Ru \cite{Paulraj, Qi2009a} (the resulting transition temperatures are shown in Table 1).  Interestingly, electron doping with Cu \cite{Canfield2009a}or hole doping with Cr \cite{Sefat2009d} does not yield superconductivity.  The large number of potential dopants together with the availability of single crystal samples has made the 122 family of compounds the topic of considerable experimental focus.

\begin{table}
\begin{minipage}[b]{0.5\linewidth}\centering
\begin{tabular}{|c|c|}
\hline
  Material & Max. T$_C$ (K)\\
  \hline
  LaFeAsO$_{1-x}$F$_x$ \cite{Kamihara2008c} & 26\\
  NdFeAsO$_{1-x}$F$_x$ \cite{Ren2008e} & 52\\
  PrFeAsO$_{1-x}$F$_x$ \cite{Ren2008d} & 52\\
  SmFeAsO$_{1-x}$F$_x$ \cite{Ren2008c} & 55\\
  CeFeAsO$_{1-x}$F$_x$ \cite{Chen2008d} & 41\\
  GdFeAsO$_{1-x}$F$_x$ \cite{Kadowaki2009} & 50\\
  TbFeAsO$_{1-x}$F$_x$ \cite{Bos2008} & 46\\
  DyFeAsO$_{1-x}$F$_x$ \cite{Bos2008} & 45\\
  Gd$_{1-x}$Th$_x$FeAsO \cite{Wang2008} & 56\\
  LaFeAsO$_{1-y}$ \cite{Ren2008b,Miyazawa2009a} & 28\\
  NdFeAsO$_{1-y}$ \cite{Ren2008b,Miyazawa2009a,Kito2008} & 53\\
  PrFeAsO$_{1-y}$ \cite{Ren2008b,Miyazawa2009a} & 48\\
  SmFeAsO$_{1-y}$ \cite{Ren2008b} & 55\\
  GdFeAsO$_{1-y}$ \cite{Yang2008c,Miyazawa2009a} & 53\\
  TbFeAsO$_{1-y}$ \cite{Miyazawa2009a} & 52\\
  DyFeAsO$_{1-y}$ \cite{Miyazawa2009a} & 52\\
  LaFe$_{1-x}$Co$_x$AsO \cite{Sefat2008a} & 14 \\
  SmFe$_{1-x}$Ni$_x$AsO \cite{Li2009k} & 10 \\
  SmFe$_{1-x}$Co$_x$AsO \cite{Qi2008a} & 15 \\
  LaFe$_{1-x}$Ir$_x$AsO \cite{Qi2009b} & 12 \\
  \hline
  \end{tabular}
\end{minipage}
\hspace{0.1cm}
\begin{minipage}[b]{0.5\linewidth}
\centering
\begin{tabular}{|c|c|}
\hline
  Material & Max. T$_C$ (K)\\
  \hline
  Ba$_{1-x}$K$_x$Fe$_2$As$_2$ \cite{Rotter2008a} & 38\\
  Ba$_{1-x}$Rb$_x$Fe$_2$As$_2$ \cite{Bukowski2009} & 23 \\
  K$_{1-x}$Sr$_x$Fe$_2$As$_2$ \cite{Sasmal2008} & 36\\  
  Cs$_{1-x}$Sr$_x$Fe$_2$As$_2$ \cite{Sasmal2008} & 37\\ 
  Ca$_{1-x}$Na$_x$Fe$_2$As$_2$ \cite{Wu2008a} & 20 \\
  Eu$_{1-x}$K$_x$Fe$_2$As$_2$ \cite{Jeevan2008a} & 32 \\
  Eu$_{1-x}$Na$_x$Fe$_2$As$_2$ \cite{Qi2008} & 35 \\ 
  Ba(Fe$_{1-x}$Co$_x$)$_2$As$_2$ \cite{Sefat2008b,Chu2009b} & 22-24 \\
  Ba(Fe$_{1-x}$Ni$_x$)$_2$As$_2$ \cite{Li2009d} & 20 \\
  Sr(Fe$_{1-x}$Ni$_x$)$_2$As$_2$ \cite{Saha2009} & 10 \\
  Ca(Fe$_{1-x}$Co$_x$)$_2$As$_2$ \cite{Kumar2009a} & 17 \\
  Ba(Fe$_{1-x}$Rh$_x$)$_2$As$_2$ \cite{Ni2009} & 24 \\
  Ba(Fe$_{1-x}$Pd$_x$)$_2$As$_2$ \cite{Ni2009} & 19 \\
  Sr(Fe$_{1-x}$Rh$_x$)$_2$As$_2$ \cite{Han2009} & 22 \\
  Sr(Fe$_{1-x}$Ir$_x$)$_2$As$_2$ \cite{Han2009} & 22 \\
  Sr(Fe$_{1-x}$Pd$_x$)$_2$As$_2$ \cite{Han2009} & 9 \\
  Ba(Fe$_{1-x}$Ru$_x$)$_2$As$_2$ \cite{Paulraj} & 21 \\
  Sr(Fe$_{1-x}$Ru$_x$)$_2$As$_2$ \cite{Qi2009a} & 13.5 \\
  LiFeAs \cite{Tapp2008, Wang2008b, Pitcher2008} & 18 \\
  Na$_{1-x}$FeAs \cite{Chu2009} & 25 \\
  Fe$_{1+y}$Se$_x$Te$_{1-x}$ \cite{Yeh2008} & 15 \\

\hline
\end{tabular}
\end{minipage}
\caption{Summary of the maximum transition temperatures at ambient pressure for various Fe-based superconductors.}
\label{tctable}
\end{table}

Superconductivity was also discovered in LiFeAs \cite{Tapp2008, Wang2008b, Pitcher2008} and Na$_{1-x}$FeAs \cite{Chu2009} (111 materials) sharing the same FeAs plane with Li or Na as the spacer as shown in Fig. \ref{structure}.  Transition temperatures of 18 K (for LiFeAs) and 12-25 K (for Na$_{1-x}$FeAs)\cite{Chu2009} have been observed dependent on the precise Na concentration.  Interestingly, superconductivity seems to appear in the 111 materials in purely stoichiometric material without chemical doping.  Finally, the Fe(Se,Te) family of compounds (11 materials) also exhibits superconductivity with a maximum transition temperature (under ambient pressure) of 15 K.  The alpha phase of FeSe is a superconductor with a transition temperature of 8 K \cite{Hsu2008}.  Structurally, the FeSe plane is very similar to the FeAs plane in the aforementioned materials (see Fig. \ref{structure}) indicating that the presence of As is not a requirement for superconductivity.  Band structure calculations\cite{Subedi2008b} suggested that FeTe may have enhanced superconducting properties.  However, pure FeTe is not a superconductor \cite{Yeh2008} but is complicated by the presence of excess Fe, \emph{i.e.} the actual chemical formula is Fe$_{1+y}$Te\cite{Bao2009}.  It has been suggested that this excess Fe is magnetic and may act as a pair breaking moment destroying superconductivity \cite{Zhang2009j}.  Nonetheless, the doped material, Fe$_{1+y}$Te$_{1-x}$Se$_x$ does exhibit an enhanced T$_C$ of 15 K with the maximum transition temperature observed for $x$ near 0.5 \cite{Yeh2008}.

The structures of the 1111, 122, 111, and 11 families of materials are shown in Fig. \ref{structure}.  The common feature is the presence of an identical FeAs (or FeSe) plane.  An interesting trend can be seen in Fig. \ref{structure} - the larger the separation between layers, the higher the observed optimal transition temperature.
Two-dimensional (2D) magnetism occurs in the regions of highest T$_C$ and, thus, may be favorable for superconductivity.
This trend led to attempts at further separating the FeAs layers and superconductivity with fairly high transition temperatures have been observed in Sr$_2$VO$_3$FeAs with a spacer of Sr$_2$VO$_3$ and T$_C$ of 37.2 K \cite{Zhu2009c}, and doped Sr$_2$Sc$_{0.4}$Ti$_{0.6}$FeAsO$_3$ with a spacer of Sr$_2$Sc$_{0.4}$Ti$_{0.6}$O$_3$ and T$_C$ onset of 45 K (although the resistivity doesn't reach zero until $\sim$ 7 K) \cite{Chen2009i}.  Although these temperatures still do not exceed the 56 K in the 1111 materials, they are quite high particularly in the case of Sr$_2$VO$_3$FeAs as this material is nominally stoichiometric.  There is hope that doping of this and related materials could lead to an increase in transition temperature.

Shortly after the discovery of superconductivity in LaFeAsO$_{1-x}$F$_x$, calculations indicated that conventional electron-phonon coupling was insufficient to explain the high transition temperatures \cite{Boeri2008}, as was later verified experimentally \cite{Christianson2008a}.  As will be explained below, a ubiquitous magnetically ordered state is present  indicating magnetism in close proximity to superconductivity leading one to naturally consider the interplay between magnetism and superconductivity in these materials.  In the following article, we will review \emph{experimental} studies of magnetism in the Fe-based compounds and its influence on superconductivity.

\section{Phase Diagrams}
\subsection{1111 Materials}
The first evidence for the importance of magnetism in the Fe-based superconductors was the concentration dependent phase diagram presented with the initial discovery of superconductivity in F-doped LaFeAsO \cite{Kamihara2008c}.  An additional phase was clearly present at low F concentration which vanished at doping levels where superconductivity appears although the exact nature of this phase was unclear.  It was soon shown that the undoped LaFeAsO parent compound exhibited spin-density wave (SDW) order below about 150 K \cite{Cruz2008, McGuire2008} consistent with a $\sqrt 2 \times \sqrt 2 \times 2$ unit cell.  Unexpectedly, LaFeAsO also exhibited a structural phase transition \cite{Cruz2008} at a temperature slightly above the magnetic ordering temperature.  The low temperature structure was originally described
by the monoclinic P112/$n$ space group \cite{Cruz2008} but it was later clarified that the correct low temperature space group is the orthorhombic $Cmma$ \cite{Nomura2008a} (note that both notations accurately describe the observed structure).  There is clear competition between magnetism and superconductivity as the magnetically ordered state is destroyed in the fluorine doped, superconducting samples \cite{Cruz2008,McGuire2008}.

\begin{figure}
\includegraphics[width=\columnwidth]{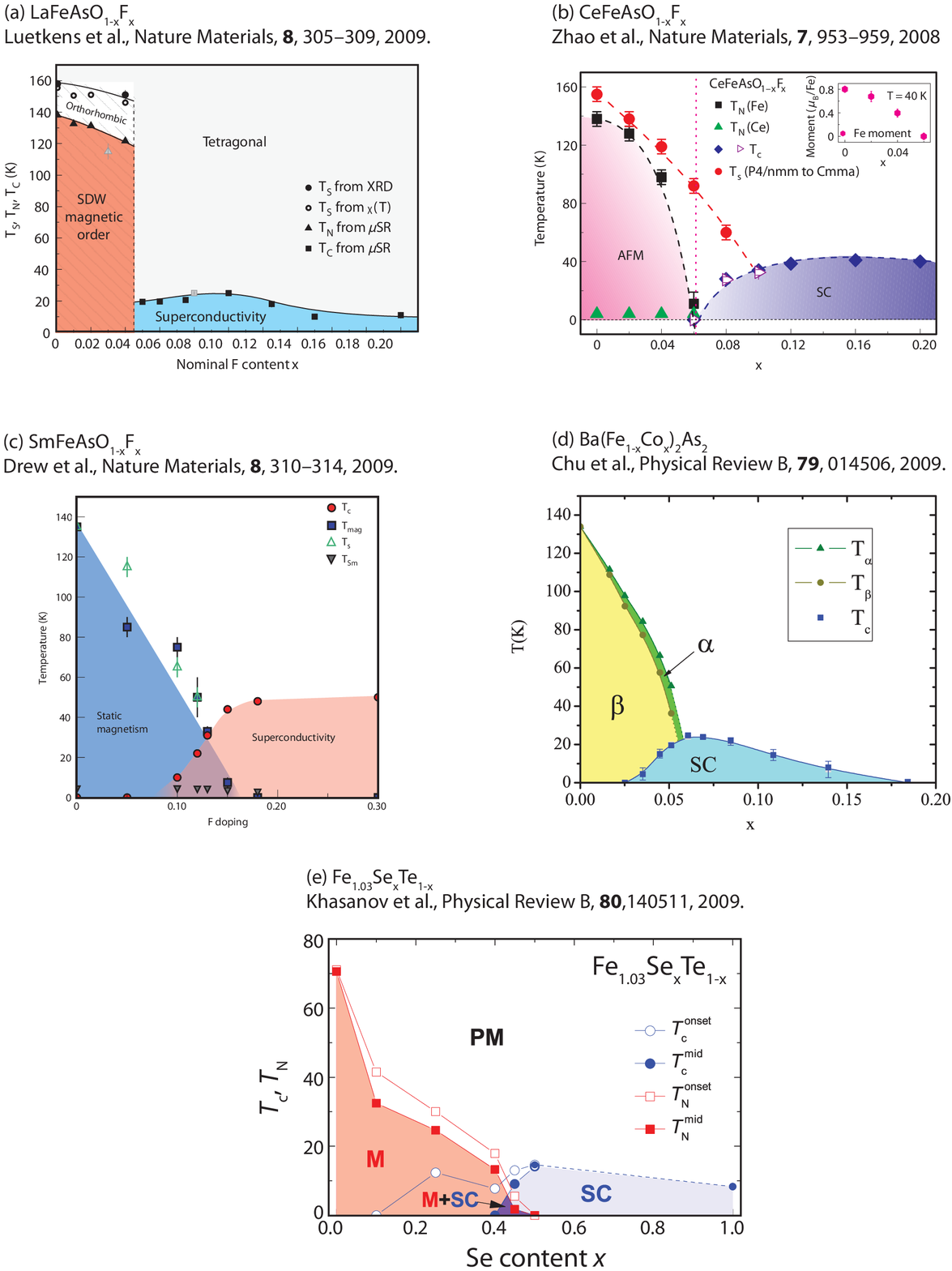}
\caption{Experimentally determined phase diagram for (a) LaFeAsO$_{1-x}$F$_x$ (Reprinted by permission from Macmillan Publishers Ltd: Nature Materials \cite{Luetkens2009}, copyright 2009), (b) CeFeAsO$_{1-x}$F$_x$  (Reprinted by permission from Macmillan Publishers Ltd: Nature Materials \cite{Zhao2008}, copyright 2008), (c) SmFeAsO$_{1-x}$F$_x$ (Reprinted by permission from Macmillan Publishers Ltd: Nature Materials \cite{Drew2009}, copyright 2009), (d) BaFe$_{2-x}$Co$_x$As$_2$ (Reprinted with permission from \cite{Chu2009b}, copyright 2009 the American Physical Society), and (e) Fe$_{1.03}$Te$_{1-x}$Se$_x$ (Reprinted with permission from \cite{Khasanov2009}, copyright 2009 the American Physical Society).}
\label{phasediagram}
\end{figure}

The phase diagram of RFeAsO$_{1-x}$F$_x$ as a function of doping has been carefully studied for R=La \cite{Luetkens2009} (Fig. \ref{phasediagram}a), Ce \cite{Zhao2008} (Fig. \ref{phasediagram}b), Pr \cite{Rotundu2009} and Sm \cite{Drew2009,Margadonna2009} (Fig. \ref{phasediagram}c).  The phase diagrams were experimentally determined using the following techniques: R=La, $\mu$SR, $^{57}$Fe Mossbauer spectroscopy and X-ray diffraction \cite{Luetkens2009}; R=Ce, neutron diffraction, resistivity and magnetization \cite{Zhao2008}; R=Pr, X-ray diffraction, resistivity and magnetization \cite{Rotundu2009}; R=Sm,  $\mu$SR \cite{Drew2009} and x-ray diffraction \cite{Margadonna2009}. For R=Nd, a partial phase diagram \cite{Chen2008c} was determined using resistivity measurements. In all cases measured, the $x$=0 parent compounds show a structural phase transition at a temperature slightly above the transition to magnetic ordering with a typical structural transition at $\sim$150 K and SDW ordering at about 140 K.  In general, doping causes a suppression of both the structural and magnetic phase transitions and as these are suppressed, superconductivity emerges.  The fundamental difference between materials with different rare earths comes in the behavior near the emergence of superconductivity.  For R=La and Pr, the structural and magnetic transitions vanish in an abrupt step-like manner as a function of doping at the onset of superconductivity \cite{Luetkens2009,Rotundu2009}, as shown in Fig. \ref{phasediagram}a for the case of R=La.  For the case of R=Ce, the magnetic transition appears to vanish continuously to very low temperatures and superconductivity emerges at a concentration where this transition has been completely suppressed \cite{Zhao2008} (see Fig. \ref{phasediagram}b).  However, the structural transition has some range of concentrations where superconductivity coexists with this phase transition \cite{Zhao2008}.  Finally, the case or R=Sm, shown in Fig. \ref{phasediagram}c, looks similar to R=Ce in that the transitions are suppressed gradually and there appears to be overlap between the structural transition and superconductivity \cite{Margadonna2009}.  However, unlike the case of Ce, the Sm phase diagram shows a region where magnetic ordering coexists with superconductivity \cite{Drew2009}.  This suggests that the destruction of long-range magnetic order is not a necessary condition for the emergence of superconductivity.

\subsection{122 Materials}

As mentioned previously, and shown in Table 1, the AFe$_2$As$_2$ family of materials has numerous doping possibilities.  The basic behavior of the superconducting materials can be described by considering the phase diagrams for Ba$_{1-x}$K$_x$Fe$_2$As$_2$ (hole doping between the FeAs planes) and BaFe$_{2-x}$Co$_x$As$_2$ (electron doping within the FeAs plane).  Both materials share the same BaFe$_2$As$_2$ parent compound.  As in the case of the 1111 parent
compounds, Ba-122 exhibits both a structural phase transition (in this case from the room temperature tetragonal $I4/mmm$ space group to the low temperature orthorhombic $Fmmm$ space group \cite{Rotter2008b,Huang2008}) and the magnetic transition to a long-range ordered, SDW state.  However, unlike the 1111 materials, both the structural and magnetic phase transitions occur at the same temperature in the Ba-122 parent compound \cite{Rotter2008b,Huang2008,Kofu2009}. Doping with either K \cite{Rotter2008,Chen2009b} or Co \cite{Chu2009b,Wang2009k,Ni2008b} causes a suppression of the structural and SDW transitions as in the 1111 materials.  For Co-doping, as $x$ increases, the two transitions no longer appear at the same temperature with the structural transition occuring first upon cooling \cite{Wang2009k} as shown in Fig. \ref{phasediagram}d. In both cases, superconductivity emerges as the SDW order is suppressed.  For K doping, the superconducting region starts for x $\sim$ 0.1 and the maximum T$_C$ of 38 K is reached for $x$ $\sim$ 0.4 \cite{Rotter2008,Chen2009b}. For Ba(Fe$_{1-x}$Co$_x$)$_2$As$_2$, superconductivity is first observed for $x$ $\sim$ 0.03 and the maximum T$_C$ of 23 K is seen for x $\sim$ 0.07 \cite{Chu2009b,Ni2008b} .  Interestingly, for both K and Co doping, there is a region of the phase diagram where the SDW state and structural transition coexist with superconductivity.

\begin{figure}
\includegraphics[width=\columnwidth]{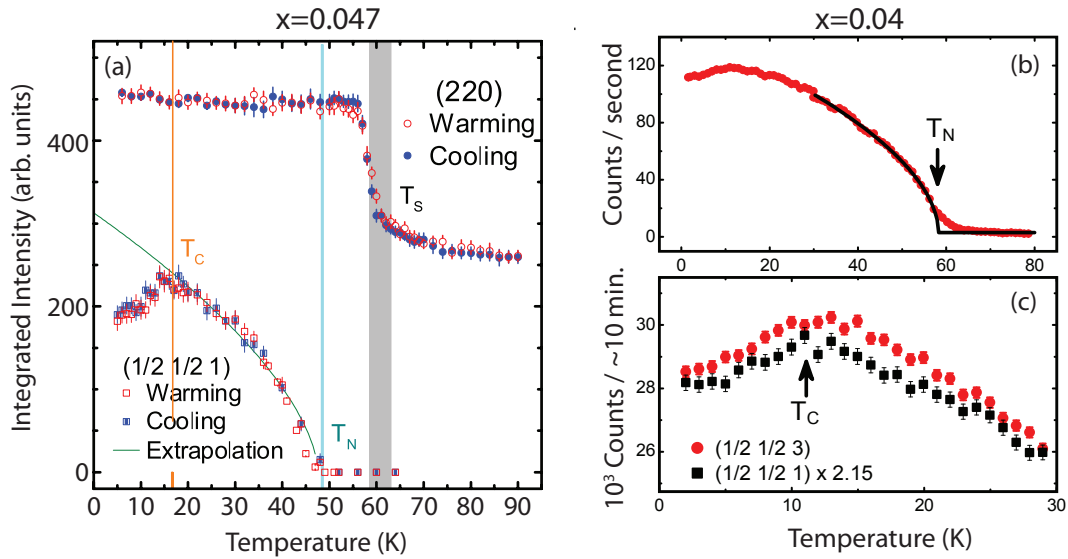}
\caption{Suppression of the magnetic Bragg peak intensity on entering the superconducting state for Ba(Fe$_{0.953}$Co$_{0.047}$)$_2$As$_2$ ($x$=0.047) (Reprinted with permission from \cite{Pratt2009}, copyright 2009 the American Physical Society) and Ba(Fe$_{0.96}$Co$_{0.04}$)$_2$As$_2$ ($x$=0.04) (Reprinted with permission from \cite{Christianson2009}, copyright 2009 the American Physical Society).}
\label{underdopeddiff}
\end{figure}

Coexistence of superconductivity and magnetism has been a recurring theme in the study of superconducting materials \cite{Chevrel1986,Muller2001,Metoki1998, Isaacs1995}.  For the doped 122 materials, the question of whether the SDW and superconducting states are microscopically coexisting or phase separated has received considerable attention experimentally.  For hole doping with K, $^{75}$As NMR \cite{Fukazawa2009}, $\mu$SR \cite{Park2009} and magnetic force microscopy \cite{Park2009} consistently indicate distinct regions which are magnetically ordered and nonmagnetic regions as expected for microscopic phase separation.  Furthermore, analysis of microstrain measured with x-ray and neutron diffraction was interpreted as being consistent with electronic phase separation \cite{Inosov2009}.  Although most measurements on the K-doped samples are consistent with a phase separation scenario, $^{57}$Fe-Mossbauer meausurements indicate a sample which is completely magnetically ordered as expected with microscopic coexistence of the SDW and superconducting states \cite{Rotter2009}.  For the case of Co doping, both $^{75}$As NMR \cite{Laplace2009} and $\mu$SR measurements \cite{Bernhard2009} indicate that all the Fe sites participate in the magnetic order as would be expected for coexistence of superconductivity and SDW order.  One $^{75}$As NMR study directly compared the cases of K and Co doping and concluded phase coexistence for Co doped samples and separation for the case of K doping \cite{Julien2009}. Finally, we note neutron diffraction measurements on Co doped samples \cite{Pratt2009,Christianson2009} showed that the magnetic Bragg peak intensity of the SDW state is suppressed on entering the superconducting state, as shown in Fig. \ref{underdopeddiff}, for x=0.04 and x=0.047.  This certainly shows a very strong interaction between the superconducting and SDW states.  It could be interpreted that this suppression is due to the same electrons participating in both the SDW and superconductivity favoring a phase coexistence scenario.  However, in a phase separation scenario, a proximity effect could cause the superconducting regions to interfere with the SDW regions causing a reduction in the SDW volume consistent with the observed Bragg peak intensity reduction.  Hence, it is difficult to make any strong conclusions about the implications of this observation for the question of phase coexistence.

\begin{figure}
\includegraphics[width=\columnwidth]{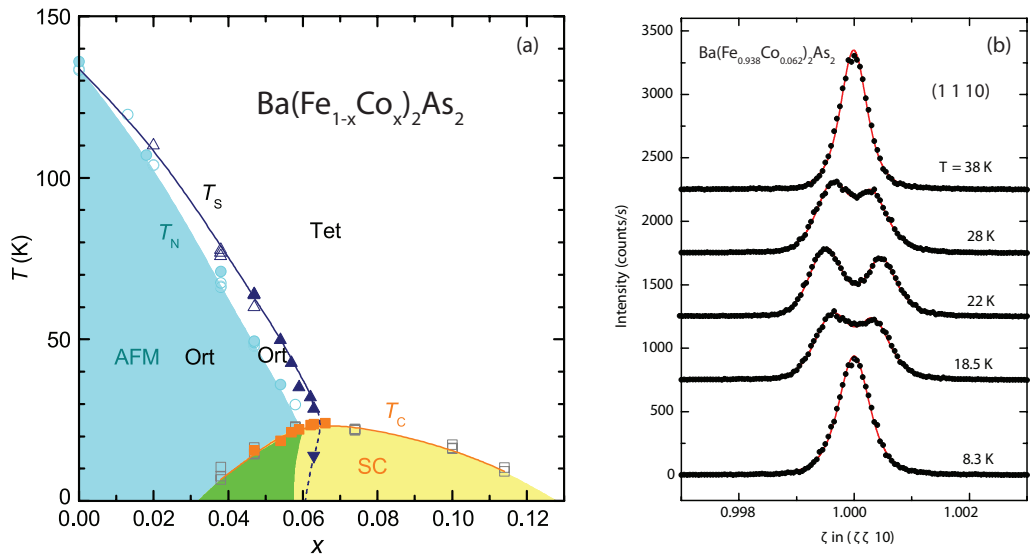}
\caption{(a) Resulting phase diagram of Ba(Fe$_{1-x}$Co$_x$)$_2$As$_2$ with inclusion of high resolution x-ray diffraction measurements in the region where superconductivity and magnetism coexist. \cite{Nandi2010} (b) Shows the reentrant nature of the structural phase transition. (Reprinted with permission from \cite{Nandi2010}, copyright 2010 the American Physical Society)}
\label{reentrant}
\end{figure}

Interestingly, the details of the phase diagram in the region where the structural and magnetic transitions cross the superconducting dome have recently been explored with high resolution x-ray diffraction \cite{Nandi2010}.  These measurements indicate that the shape of the line in the $x$-T phase diagram representing the tetragonal to orthorhombic transition changes on entering the superconducting state and bends to lower values of $x$ \cite{Nandi2010} as shown in Fig. \ref{reentrant}a.  As such, clear reentrant behavior is seen in a crystal of Ba(Fe$_{0.938}$Co$_{0.062}$)$_2$As$_2$ where the system transforms from tetragonal to orthorhombic and back to tetragonal on cooling \cite{Nandi2010} (see Fig. \ref{reentrant}b).  This shows a strong interaction between the structural transition and superconductivity and it was proposed that the interaction was actually one between magnetism and superconductivity with the influence on the structural transition resulting from magneto-elastic coupling \cite{Nandi2010}.

\subsection{11 Materials}
Finally, we discuss the phase diagram of the FeSe$_x$Te$_{1-x}$ family of materials.  As mentioned previously, these materials form with excess Fe with the largest amount of extra Fe observed near the Te-rich side of the phase diagram.  Initial measurements of the Fe$_{1+y}$Te$_{1-x}$Se$_x$ \cite{Yeh2008} family of compounds showed superconductivity with T$_C$ as high as 15 K for x$\sim$0.5 existing for all values of $x$ except very near $x$=0 where superconductivity is destroyed. This suggests a different phase diagram from other Fe based superconductors.  However, single crystal specific heat measurements indicated bulk superconductivity only for concentrations near $x$=0.5 \cite{Sales2009}.  With this in mind, the phase diagram was reinvestigated and indicated magnetic order for small $x$ which coexists with superconductivity over a range of concentrations \cite{Khasanov2009} (see Fig. \ref{phasediagram}e) in a manner very similar to the doped 122 materials and SmFeAsO$_{1-x}$F$_x$.  As mentioned previously, materials with low Se concentrations have a tendency to form with excess Fe.  Measurements of the phase diagram with samples intentionally grown with Fe$_{1.1}$ \cite{Paulose2009} show an additional spin glass phase which coexists with superconductivity over much of the measured concentration range.  This shows the sensitivity of these materials to stoichiometry and, in particular, the amount of excess Fe present.

Although, as discussed above, there are some differences in the concentration dependent phase diagrams of various Fe-based superconductors, inspection of Fig. \ref{phasediagram} shows that there are some common features.  All materials exhibit a SDW state at low concentrations and this state is suppressed with doping allowing for the emergence of superconductivity.  This shows strong similarity to the generic cuprate phase diagram and is evidence for the interplay of magnetism and superconductivity in the Fe-based materials.

\section{Magnetic Order}

The parent compounds of both the 1111 and 122 materials are metals which exhibit SDW order.  The high temperature (T $>$ T$_N$) paramagnetic state is characterized by magnetic susceptibility with an unusual linear temperature dependence ($\chi$ $\propto$ T) \cite{Klingeler2010,Wu2008a,Wang2009j,Yan2008,Zhang2009b,Sales2009a}. This behavior is neither Pauli nor Curie-Weiss like and is reminiscent of the (T$>$T$_{SDW}$) behavior of metallic Cr \cite{McGuire1952}.  In the following section, we will provide an overview of the magnetic order which evolves out of this unusual paramagnetic state in the 1111, 122, and 11 family of materials.  Examination of the magnetic ordering can shed light on the magnetic interactions and the nature (local moment or itinerant) of the magnetism in these materials.

\subsection{1111 Materials}
As discussed above, it is well established that the undoped parent compounds exhibit some form of antiferromagnetic long range order.  This was first observed in LaFeAsO where the magnetic structure was characterized by the ordering wavevector (1/2 1/2 1/2)$_T$ = (1 0 1/2)$_O$ (where the subscripts $T$ and $O$ refer to the tetragonal and orthorhombic structures, respectively) and the low temperature ordered magnetic moment was 0.36 $\mu$B \cite{Cruz2008}.  At this point, we note that the ordering wavevector in the orthorhombic cell (\emph{i.e.} (1 0 1/2)$_O$) differs from the wavevector listed in \cite{Cruz2008} as the later wavevector is relative to the unit cell of the magnetic structure where the unit cell is doubled along the c-axis.  The observed wavevector is consistent with a magnetic unit cell of size $\sqrt{2}$a $\times$ $\sqrt{2}$a $\times$ 2c relative to the tetragonal cell.  This ordering is consistent with stripe-like antiferromagnetic order with ferromagnetically coupled chains along the tetragonal (110) direction coupled antiferromagnetically along the in-plane perpendicular direction. The doubling of the unit cell along the $c$-axis indicates antiferromagnetic interactions between neighboring planes. The magnetic moment direction could not be uniquely determined in this measurement but the observed intensity is consistent with moments lying in the a-b plane.  The magnetic moment observed is much smaller than the 2.2 $\mu$B moment observed in metallic Fe.  Measurements of LaFeAsO$_{1-x}$F$_x$ shows that the magnetic moment is rather independent of concentration for $x<$0.03 and is zero for $x>$0.05 \cite{Huang2008a}.  More concentration points are required to determine how abruptly the magnetic moment vanishes with fluorine concentrations between 3 and 5 $\%$.

The nature of the ordered state in these materials has been a topic of considerable study.  The calculated Fermi surface for LaFeAsO consists of electron cylinders near the M point and hole cylinders and a 3D hole pocket around the $\Gamma$ point \cite{Singh2008a}.  Further investigations indicated good nesting of these components separated by the 2D
 wavevector (1/2 1/2)$_T$ consistent with the observed magnetic structure \cite{Mazin2008a,Dong2008}.  This led to the suggestion that the observed antiferromagnetic state is a SDW induced by Fermi surface nesting \cite{Dong2008}.  In addition to this Fermi surface nesting scenario, it has been proposed that near-neighbor and next-near-neighbor interactions between local Fe moments are both antiferromagnetic and of comparable strength leading to magnetic frustration \cite{Yildirim2008,Ma2009a,Si2008}.  In addition to describing the observed magnetic structure, this scenario can also provide an explanation for the structural phase transition as the lattice distortion relieves the magnetic frustration \cite{Yildirim2008,Ma2009a}.  These frustration effects have also been used to explain the small ordered moment \cite{Yildirim2008,Si2008}.  Starting with a local moment Hamiltonian consistent with those discussed previously \cite{Yildirim2008,Ma2009a,Si2008}, it was suggested that the structural transition is actually a transition to a ``nematic" ordered phase which will occur at a higher temperature than the SDW transition \cite{Fang2008}.  In addition to the view that the magnetic order is driven exclusively by either Fermi surface nesting or local moment superexchange interactions, an alternate approach based on analysis of DFT calculations included aspects of both \cite{Johannes2009}.  This work concluded that the moments were largely local in nature but the interactions were relatively long-ranged itinerant interactions as opposed to superexchange and both the low temperature magnetic order and structural distortions were explained \cite{Johannes2009}. Finally, it was recently proposed that both the magnetic and structural transitions are driven by orbital physics and that the structural transition is, in fact, a ferro-orbital ordering transition \cite{Lee2009f}.  This model explains the coupling of the structural and magnetic transitions and is consistent with the rather large ordering temperature \cite{Lee2009f}.

Changes of the ordered magnetic structure with different rare earth elements (RFeAsO) have been extensively studied with neutron diffraction as well as local probe methods.  The ordering wavevector of (1/2 1/2 1/2)$_T$ observed for R=La \cite{Cruz2008} is also observed for R=Nd \cite{Chen2008h}.  However, for R=Ce \cite{Zhao2008} and R=Pr \cite{Zhao2008a} the ordering is described by the wavevector (1/2 1/2 0) suggesting ferromagnetic coupling between planes.  This suggests rather weak interplane coupling which is strongly influenced by the rare earth ion and the associated induced structural changes.  Unfortunately, for the case of R=Sm, the high absorption cross-section for Sm makes neutron scattering measurements very difficult.  Neutron scattering measurements on SmFeAsO were performed \cite{Ryan2009} but could only explore the low temperature ordering of the Sm moments as will be discussed below.

\begin{figure}
\centering
\includegraphics[width=0.85\columnwidth]{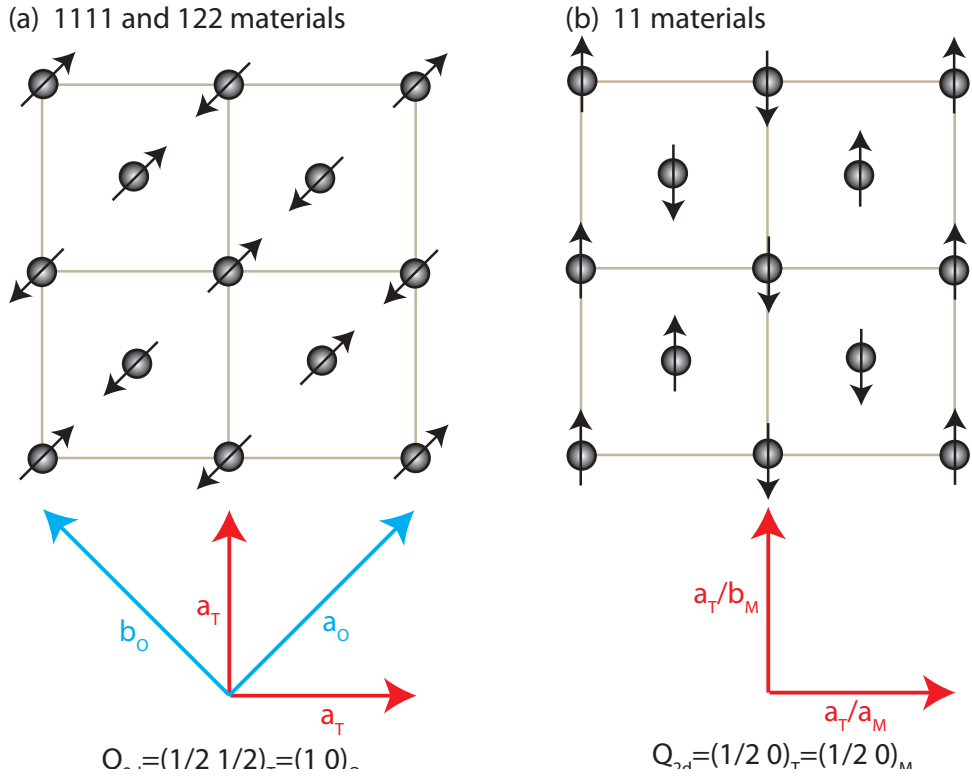}
\caption{(a) In plane magnetic structure for the 1111 and 122 parent compounds.  The ordering wavevector in these compounds is (1/2 1/2 L)$_T$=(1 0 L)$_O$.  For the 1111 materials, the stacking of neighboring plane along the $c$-axis is either ferromagnetic or antiferromagnetic depending on the rare earth element (see Table 2).  For the 122 materials, the stacking is antiferromagnetic along the c-axis resulting in odd-integer L as the unit cell contains two FeAs layers. (b) Magnetic structure for 11 materials (Fe$_{1+x}$Te) in the limit of smaller $x$ where the low temperature nuclear structure is monoclinic.  The ordering wavevector is (1/2 0 1/2) and is the same in both the high and low temperature phases.}
\label{magstructure}
\end{figure}

\begin{figure}
\includegraphics[width=\columnwidth]{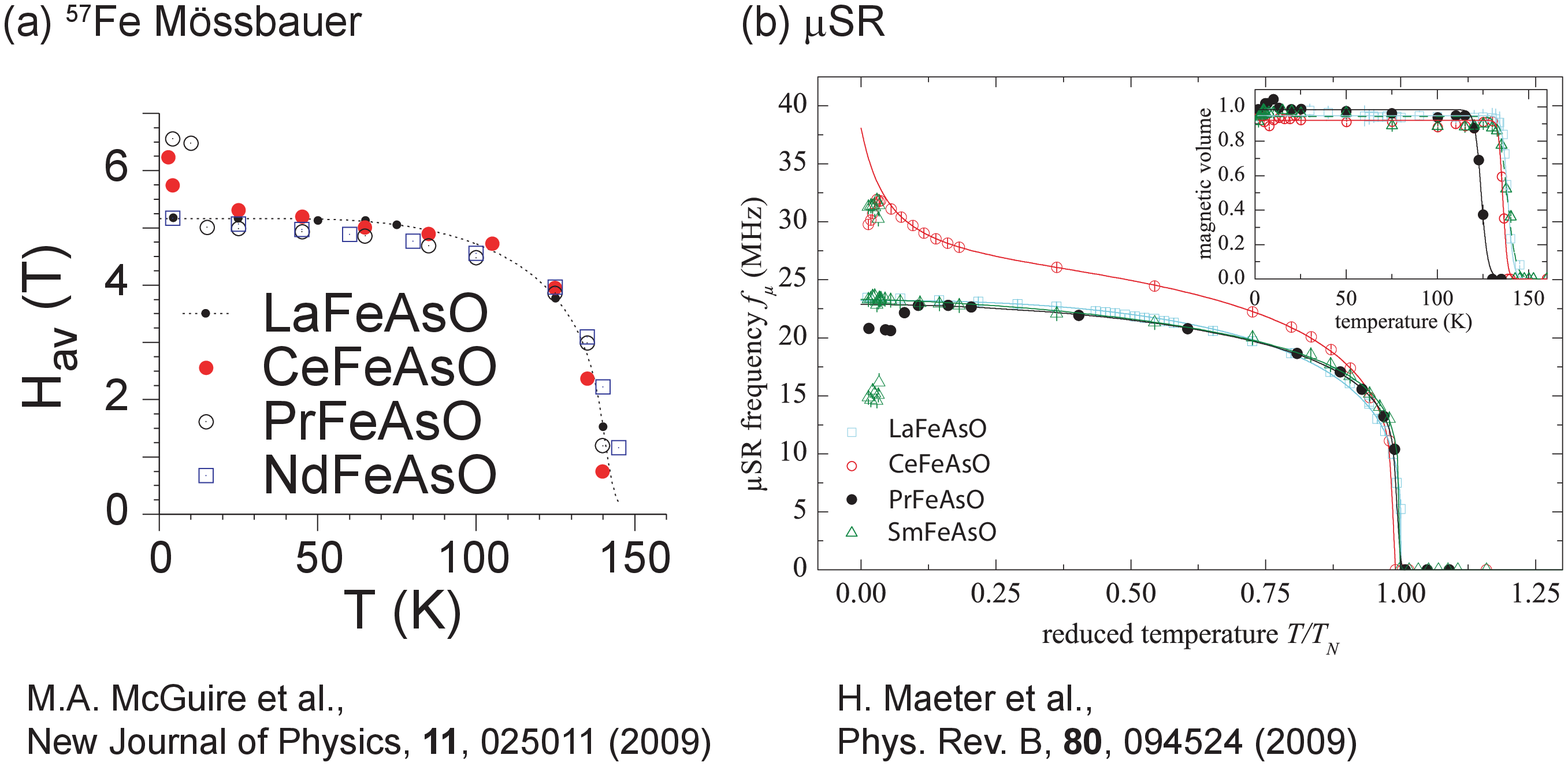}
\caption{Measurement of the ordered moment from (a) $^{57}$Fe M\"{o}ssbauer (Reprinted with permission from \cite{McGuire2009}, copyright 2009 IOP Publishing) and (b) $\mu$SR measurements (Reprinted with permission from \cite{Maeter2009}, copyright 2009 the American Physical Society).  The M\"{o}ssbauer measurements indicate a moment independent of rare earth element while $\mu$SR measurements show a noticeably different moment for the case of CeFeAsO. Note that the $\mu$SR measurements on CeFeAsO do not appear to saturate at low temperature consistent with a temperature dependent polarization of Ce.}
\label{momentcompare}
\end{figure}

The size of the ordered moment as a function of R has been a topic of considerable interest.  Neutron scattering on R=Pr indicates a moment of 0.34 $\mu$B \cite{Kimber2008} identical to that observed for R=La \cite{Cruz2008} (a moment of 0.48 $\mu$B \cite{Zhao2008a} was independently observed but this was measured below the Pr ordering temperature).  The moment for R=Nd appears smaller and initially, Fe ordering wasn't observed \cite{Qiu2008} with an upper bound on the ordered moment placed at 0.17 $\mu$B. However, later measurements clearly indicated Fe ordering with an ordered moment of 0.25 $\mu$B \cite{Chen2008h}, the smallest of any of the rare earths.  A particularly interesting case is that of Ce where neutron scattering indicated a much larger magnetic moment of 0.8 $\mu$B \cite{Zhao2008} more than twice the size of any other rare earth.  Thus, on the basis of these neutron diffraction results, the Fe moment size varies considerably with rare earth element. However, a contradictory picture is obtained from $^{57}$Fe M\"{o}ssbauer mesurements.  Such measurements for R=La indicate an internal magnetic field of 4.86 T \cite{Klauss2008}, 5.19 T \cite{McGuire2008}, and 5.3 T \cite{Kitao2008}. For the other rare earths, the internal field was measured to be 5.2 T \cite{Sanchez2009} and 5.3 T \cite{McGuire2009} for R=Nd, 5.06 T \cite{McGuire2009} for R=Ce, and 4.99 T \cite{McGuire2009} for R=Pr.  Averaging for multiple values on the same material and using the conversion that 15 T internal field corresponds to 1 $\mu$B \cite{McGuire2008,McGuire2009}, yields an ordered Fe moment to be 0.34 $\mu$B for R=La, 0.35 $\mu$B for R=Nd, 0.34 for R=Ce, and 0.33 for R=Pr.  This suggests an ordered Fe moment size which is independent of rare earth ion, as shown in Fig. \ref{momentcompare}a where the results for R=La, Ce, Pr and Nd are superposed, in apparent contradiction to the neutron results.  Zero-field $\mu$SR measurements indicate a spontaneous muon spin precession frequency below T$_N$ for all rare earth parent compounds consistent with long-range antiferromagnetic order \cite{Maeter2009}.  The size of the magnetic moment is reflected by the saturation frequency which is about 23 MHz for R=La,Pr,Nd, and Sm \cite{Klauss2008,Maeter2009,Aczel2008,Carlo2009,Drew2009}.  Interestingly, measurements for R=Ce indicates a significantly higher saturation frequency \cite{Maeter2009}, as can be seen in Fig. \ref{momentcompare}b, suggesting a larger ordered moment consistent with the neutron diffraction measurements and inconsistent with the M\"{o}ssbauer results.  One explanation for this discrepancy is polarization of the Ce sublattice by the ordered Fe moments \cite{Maeter2009}.  Such a polarization would affect the $\mu$SR measurements by modifying the local magnetic field at the muon site and would affect the neutron Bragg reflections under the assumption that the polarized Ce moments exhibit the same periodicity as the ordered Fe moments.  The M\"{o}ssbauer measurements are expected to be less affected as the influence of the Ce moments on the Fe hyperfine field should be small \cite{Maeter2009}.

For the parent compounds with magnetic rare earth ions (i.e. Pr, Ce, Nd, and Sm), the rare earth moments order at low temperatures.  The Pr moments in PrFeAsO order below 14 K \cite{Zhao2008a,Kimber2008} with a fairly complex ordered structure with Pr spins along the c-axis \cite{Zhao2008a}.  There is coupling between the Pr and Fe moments and the ordered moments at 5 K were reported to be 0.84 $\mu$B for Pr and 0.48 $\mu$B for Fe \cite{Zhao2008a} (an independent measurement indicated moments at 1.4 K of 0.83 $\mu$B for Pr and 0.53 $\mu$B for Fe \cite{Kimber2008}).  Note that the Fe moment is enhanced from the value of 0.35 $\mu$B observed for temperatures above the Pr ordering temperature \cite{Kimber2008}.  Ce moments in CeFeAsO order below $\sim$4 K with moments lying primarily in the $a$-$b$ plane \cite{Zhao2008}.  As in the case of Pr, significant coupling between the Fe and Ce moments is observed with low temperature ordered moments of 0.83 $\mu$B and 0.94 $\mu$B for Ce and Fe respectively which can be compared to the Fe moment of 0.8 $\mu$B at 40 K \cite{Zhao2008}.  Nd spins in NdFeAsO order below 2 K and form a collinear arrangement with antiferromagnetic coupling along the orthorhombic $b$ axis \cite{Qiu2008}.  The ordered moments below 2 K were found to be 1.55 $\mu$B for Nd and 0.9 $\mu$B for Fe indicating a strong enhancement when compared to the Fe ordered moment of 0.25 $\mu$B observed for temperatures between the Fe and Pr ordering temperatures \cite{Chen2008h}.  Finally, despite the large absorption cross-section of Sm, low temperature measurements on SmFeAsO indicated Sm order at 1.6 K \cite{Ryan2009}.  The determined Sm spin structure is quite different than the other rare earths in that ferromagnetic sheets of Sm moments are stacked antiferromagnetically along the c-axis and the ordered Sm moment is 0.6 $\mu$B \cite{Ryan2009}.  In contrast, the cases of Pr, Ce, and Nd have rare earth moments coupled antiferromagnetically along the $b$ axis despite differences in the moment direction \cite{Zhao2008a,Zhao2008,Qiu2008}.  Note that for all measured cases, the Fe ordering arrangement exhibits ferromagnetic coupling along the orthorhombic $b$ axis with an antiferromagnetic arrangement along the $a$-axis.

\begin{table}
\begin{tabular}{|c|c|c|c|c|c|}
  \hline
  Material & T$_N$ (K) & Wavevector & Moment  & Technique & Ref. \\
  \hline
  LaFeAsO & 137 & (1 0 $\frac{1}{2}$)$_O$ / ($\frac{1}{2}$ $\frac{1}{2}$ $\frac{1}{2}$)$_T$ & 0.36 $\mu$B& neutrons & \cite{Cruz2008} \\
     & 138 & & 4.86 T / 23 MHz & M\"{o}ssbauer / $\mu$SR & \cite{Klauss2008}\\
     & 145 & & 5.19 T & M\"{o}ssbauer & \cite{McGuire2008} \\
     & 140 & & 5.3 T & M\"{o}ssbauer & \cite{Kitao2008} \\
  NdFeAsO & 141 & (1 0 $\frac{1}{2}$)$_O$ / ($\frac{1}{2}$ $\frac{1}{2}$ $\frac{1}{2}$)$_T$ & 0.25 $\mu$B& neutrons & \cite{Chen2008h} \\
       & 141 & & 5.3 T & M\"{o}ssbauer & \cite{McGuire2009} \\
       & 135 & & 23 MHz & $\mu$SR & \cite{Aczel2008} \\
  PrFeAsO & 136 & (1 0 0)$_O$ / ($\frac{1}{2}$ $\frac{1}{2}$ 0)$_T$ & 0.35 $\mu$B& neutrons & \cite{Kimber2008} \\
       & 139 & & 4.99 T & M\"{o}ssbauer & \cite{McGuire2009} \\
       & 123 & & 23 MHz & $\mu$SR & \cite{Maeter2009} \\
  CeFeAsO & 140 & (1 0 0)$_O$ / ($\frac{1}{2}$ $\frac{1}{2}$ 0)$_T$ & 0.8 $\mu$B& neutrons & \cite{Zhao2008} \\
       & 136 & & 5.06 T & M\"{o}ssbauer & \cite{McGuire2009} \\
       & 137 & & $\sim$26 MHz & $\mu$SR & \cite{Maeter2009} \\
  SmFeAsO & 135 & & 23.6 MHz & $\mu$SR & \cite{Drew2009} \\
  BaFe$_2$As$_2$ & 90$^*$ & (1 0 1)$_O$ / ($\frac{1}{2}$ $\frac{1}{2}$ 1)$_T$ & 0.99 $\mu$B & neutrons & \cite{Su2009} \\
       & 143 & (1 0 1)$_O$ / ($\frac{1}{2}$ $\frac{1}{2}$ 1)$_T$ & 0.87 $\mu$B & neutrons & \cite{Huang2008} \\
       & 140 &  & 28.8 MHz & $\mu$SR & \cite{Aczel2008} \\
       & 140 &  & 5.47 T & M\"{o}ssbauer & \cite{Aczel2008} \\
  SrFe$_2$As$_2$ & 220 & (1 0 1)$_O$ / ($\frac{1}{2}$ $\frac{1}{2}$ 1)$_T$ & 0.94 $\mu$B & neutrons & \cite{Zhao2008b} \\
       & 205 & (1 0 1)$_O$ / ($\frac{1}{2}$ $\frac{1}{2}$ 1)$_T$ & 1.01 $\mu$B & neutrons & \cite{Kaneko2008} \\
       & 205 &  & 44 MHz & $\mu$SR & \cite{Jesche2008} \\
       & 205 &  & 8.91 T & M\"{o}ssbauer & \cite{Tegel2008b} \\
  CaFe$_2$As$_2$ & 173 & (1 0 1)$_O$ / ($\frac{1}{2}$ $\frac{1}{2}$ 1)$_T$ & 0.8 $\mu$B & neutrons & \cite{Goldman2008} \\
  EuFe$_2$As$_2$ & 200 &  & 8.5 T & M\"{o}ssbauer & \cite{Raffius1993} \\

  Fe$_{1.076}$Te & 75 & ($\frac{1}{2}$ 0 $\frac{1}{2}$) & 2.03 $\mu$B & neutrons & \cite{Bao2009} \\
  Fe$_{1.141}$Te & 63 & (0.38 0 $\frac{1}{2}$) & 1.96 $\mu$B & neutrons & \cite{Bao2009} \\
  Fe$_{1.068}$Te & 67 & ($\frac{1}{2}$ 0 $\frac{1}{2}$) & 2.25 $\mu$B & neutrons & \cite{Li2009g} \\

  Na$_{1-\delta}$FeAs & 37 & (1 0 $\frac{1}{2}$)$_O$ / ($\frac{1}{2}$ $\frac{1}{2}$ $\frac{1}{2}$)$_T$ & 0.09 $\mu$B & neutrons & \cite{Li2009f} \\

  \hline
\end{tabular}
\caption{Magnetic structure parameters for Fe order.  The ordered moment for $^{57}$FeM\"{o}ssbauer measurements is shown in units of internal magnetic field and can be converted to $\mu$B such that 15 T internal field corresponds to 1 $\mu$B \cite{McGuire2008,McGuire2009}.  For $\mu$SR measurements, the ordered moment is presented as a saturation frequency in MHz.  Note that we characterize the magnetic structure in this table by an ordering wavevector relative to the respective (tetragonal or orthorhombic) nuclear cell.  This differs from the wavevector listed in some references (for instance, \cite{Lynn2009}) where the characteristic wavevector is relative to a unit cell expanded to fully include the magnetic structure forcing all indices to be integer.}
\end{table}

\subsection{122 Materials}

The temperature dependent structure of the AFe$_2$As$_2$ parent compounds has been carefully studied for BaFe$_2$As$_2$ \cite{Rotter2008b,Huang2008}, SrFe$_2$As$_2$ \cite{Yan2008,Tegel2008b,Jesche2008}, CaFe$_2$As$_2$ \cite{Ni2008a} and EuFe$_2$As$_2$ \cite{Tegel2008b}.  In all cases the room temperature tetragonal space group is $I4/mmm$ and the materials transform to a low temperature orthorhombic $Fmmm$ space group with a 45$^\circ$ rotated cell in the $a$-$b$ plane.  The room temperature $I4/mmm$ space group is different than the $P4/nmm$ space group of the 1111 materials in that it contains two FeAs layers per unit cell.
In contrast to the 1111 materials, the 122 parent compounds exhibit a structural and magnetic transition at the same temperature as shown in measurements on BaFe$_2$As$_2$ \cite{Huang2008,Su2009}, SrFe$_2$As$_2$ \cite{Zhao2008b,Kaneko2008,Jesche2008,Li2009a}, and CaFe$_2$As$_2$ \cite{Goldman2008}.  Despite some initial contradicting reports, there is now consensus that the magnetic structure is the same in all measured parent compounds.  Neutron diffraction measurements on BaFe$_2$As$_2$ \cite{Huang2008,Su2009,Kofu2009}, SrFe$_2$As$_2$ \cite{Zhao2008b,Kaneko2008}, and CaFe$_2$As$_2$ \cite{Goldman2008} all indicate a magnetic structure characterized by a (1 0 1)$_O$ wavevector (where the orthorhombic cell is defined such that $c>a>b$) with moments oriented along the $a$ axis arranged antiferromagnetically along $a$ and ferromagnetically along $b$.  Neighboring layers are stacked antiparallel to one another along the $c$-axis.  Note that in tetragonal notation, this wavevector corresponds to (1/2 1/2 1)$_T$. The odd-integer value of L in the ordering wavevector is a consequence of the antiferromagnetic stacking along the $c$-axis together with the presence of two layers in a tetragonal unit cell.

Neutron diffraction experiments on the 122 materials find larger ordered magnetic moments than in the 1111 materials which are fairly consistent for different members of the AFe$_2$As$_2$ family with 0.99 $\mu$B \cite{Su2009} observed in single crystal measurements on BaFe$_2$As$_2$ (grown with Sn flux), 0.87 $\mu$B \cite{Huang2008} in powder measurements on BaFe$_2$As$_2$, 0.94 $\mu$B in single crystals of SrFe$_2$As$_2$, 1.01 $\mu$B \cite{Kaneko2008} in powder measurements on SrFe$_2$As$_2$ and 0.8 $\mu$B in single crystals of CaFe$_2$As$_2$.  This consistency of the ordered moment occurs despite a large variation in transition temperatures ranging from 90 K in BaFe$_2$As$_2$ grown with Sn flux \cite{Su2009} to 220 K in crystals of SrFe$_2$As$_2$ \cite{Zhao2008b}.  On the other hand, $^{57}$Fe M\"{o}ssbauer results indicate an internal field of 5.47 T for BaFe$_2$As$_2$ \cite{Aczel2008} and much different values of 8.91 T for SrFe$_2$As$_2$ \cite{Tegel2008b} and 8.5 T for EuFe$_2$As$_2$ \cite{Raffius1993}.
Interestingly, the M\"{o}ssbauer internal field shows a similar trend as the magnetic ordering temperature with internal fields of 5.47:8.91:8.5 and transition temperatures of 140:205:200 for Ba:Sr:Eu suggesting that the increased transition temperature is the result of enhanced magnetism in the cases of Sr and Eu.  The consistency of magnetic moments from neutron diffraction and inconsistency of the internal fields from M\"{o}ssbauer measurements in the 122 samples is precisely the opposite of observations on the 1111 materials where the neutron moments were quite different for different rare earths while the M\"{o}ssbauer fields were the same.  Although the reason for this difference is unclear, it was pointed out that for M\"{o}ssbauer measurements the proportionality between the hyperfine field and the ordered moment is questionable particularly in light of a potentially unquenched orbital moment \cite{Bonville2010}.

\begin{figure}
\includegraphics[width=\columnwidth]{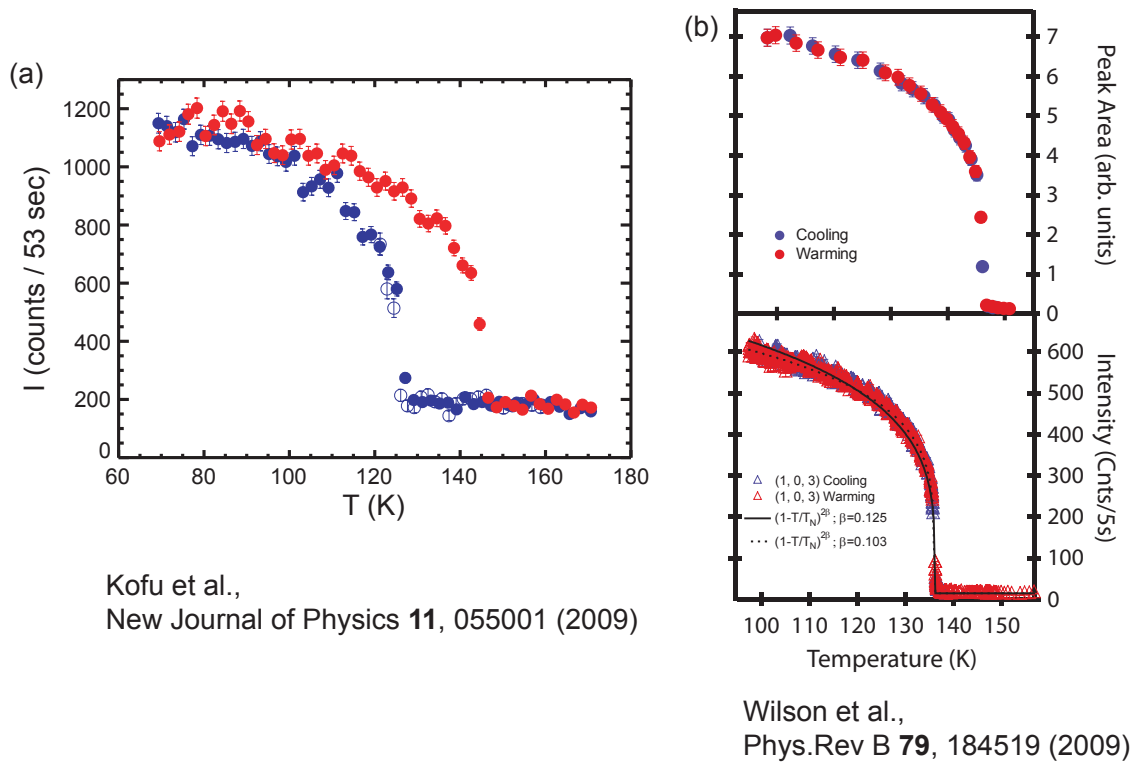}
\caption{Measurements of the Bragg peak intensity on two separate single crystal samples on cooling and warming. The heating / cooling rate in (a) was 2.3 K /min (solid points) (Reprinted with permission from \cite{Kofu2009}, copyright 2009 IOP Publishing) while that in (b) was very slow ($<$ 0.05 K /min average rate) (Reprinted with permission from \cite{Wilson2009}, copyright 2009 the American Physical Society).  Strong hysteresis is seen in \cite{Kofu2009} while no hysteresis was observed in \cite{Wilson2009}.  It seems likely that the large difference in hysteresis cannot be fully explained by the rate of temperature change.  It has been suggested \cite{Wilson2009} that difference in sample quality or perhaps external strain due to sample mounting may explain the difference between the hysteresis in various measurements.}
\label{hysteresiscompare}
\end{figure}

The phase transition in the AFe$_2$As$_2$ compounds is rather abrupt and the continuous or discontinuous nature of the phase transition in these compounds has been a topic of considerable study.  An abrupt change in the order parameter in of itself is not evidence of a first-order transition and may simply be the result of a small critical exponent, $\beta$, in a continuous transition.  Hysteresis has been observed in neutron scattering \cite{Kofu2009,Li2009a,Goldman2008}, NMR \cite{Kitagawa2008,Kitagawa2009} and specific heat measurements \cite{Krellner2009} suggesting a first-order phase transition.  This hysteresis, however, is found to be sample and thermal history dependent.  As an example, neutron diffraction measurements on a single crystal BaFe$_2$As$_2$ sample show a large hysteresis ($\sim$20 K) upon rather rapid warming and cooling (minimum of 2.3 K / minute) \cite{Kofu2009} (as shown in Fig. \ref{hysteresiscompare}a) while neutron powder diffraction measurements on BaFe$_2$As$_2$ under conditions of slower warming and cooling (0.25 K / min) shows a smaller hysteresis of less than 10 K \cite{Huang2008}.  Careful neutron diffraction order parameter measurements also on a single crystal of BaFe$_2$As$_2$ with a very slow rate of cooling and warming ($<$0.05 K /min overall rate) indicate no measurable hysteresis \cite{Wilson2009} as shown in Fig. \ref{hysteresiscompare}b.  The difference in measured hysteresis shown in Fig. \ref{hysteresiscompare} is quite striking and it seems unlikely that such differences can be attributed purely to the rate of temperature change.  This suggests that the magnitude of the hysteresis is sample dependent making it difficult to extract the true nature of the phase transition.  In addition to hysteresis, further evidence for the first order nature of the transition was shown in the form of phase coexistence over a finite temperature range observed in neutron diffraction experiments on a single crystal sample of SrFe$_2$As$_2$ \cite{Li2009a}.

On the other hand, there is some evidence for a continuous phase transition.  We note that the low temperature orthorhombic $Fmmm$ space group is a subgroup of the high temperature tetragonal $I4/mmm$ space group as would be expected for a continuous structural phase transition \cite{Tegel2008b,Rotter2008b}.  Several measurements including neutron diffraction \cite{Wilson2009} (Fig. \ref{hysteresiscompare}b) and x-ray diffraction \cite{Jesche2008} showed no measurable hysteresis.  Furthermore, the order parameter can be well parameterized by a power law fit albeit with a very small critical exponent consistent with 2D Ising behavior \cite{Tegel2008b,Wilson2009}.  It has been suggested \cite{Wilson2009} that the disparate hysteretic behavior is a consequence of high sensitivity to strain.  Such strain may be imposed internally by the inclusion of sample dependent impurities or externally due to the mounting of the crystal and may be sufficient to modify the nature of the phase transition. Indeed, BaFe$_2$As$_2$ has been known to be quite sensitive to the inclusion of impurities where samples grown in the presence of Sn flux (see, for instance \cite{Su2009}) have a structural / magnetic phase transition at $\sim$90 K while self-flux grown samples have a very different transition temperature of $\sim$ 140 K.  This difference has been attributed to the inclusion of Sn impurities.

Finally we discuss the evolution of the magnetic structure with doping.  Most measurements have focussed on Ba(Fe$_{1-x}$Co$_x$)$_2$As$_2$ as the crystals are considered to be homogeneous.  Neutron scattering measurements have indicated a magnetic structure characterized by the same (1 0 1)$_O$ wavevector as the BaFe$_2$As$_2$ parent compound for concentrations of $x$=0.04 \cite{Christianson2009} and $x$=0.047 \cite{Pratt2009}.  However, NMR measurements on a sample with $x$=0.06 indicate a distribution of internal fields \cite{Laplace2009} and $^{57}$ M\"{o}ssbauer measurements on single crystals with $x$ as high as 0.045 indicate a distribution of hyperfine fields \cite{Bonville2010}.  Both observations \cite{Laplace2009,Bonville2010}, as well as NMR measurements on other underdoped samples \cite{Ning2009a}, were taken as evidence that the SDW order evolves from commensurate for the parent compound to incommensurate in the presence of Co doping.  Modeling the NMR lineshape yields the prediction of a small incommensuration with magnitude $\varepsilon$ $\sim$ 0.04 \cite{Laplace2009}.  The particular case considered involved a (1-$\varepsilon$,0,L)$_O$) wavevector \cite{Laplace2009}.  We note that for the neutron measurements reported \cite{Pratt2009,Christianson2009} the resolution at the orthorhombic (1 0 1) wavevector is less than 0.02 r.l.u. along $h$ which would have easily allowed for measurements of an incommensuration of the value estimated above.  However, an incommensuration significantly smaller than this resolution is certainly possible.  Furthermore, the resolution along the orthorhombic $k$ direction (the vertical direction experimentally) is very coarse and an incommensuration along this direction would be difficult to detect in these experiments.  High resolution neutron diffraction experiments are needed to resolve this issue.

\subsection{11 Materials}

The structural and magnetic properties of the 11 family of compounds is complicated by extreme sensitivity to stoichiometry and the presence of excess Fe.  As an example, nearly stoichiometric FeSe$_{0.97}$ crystallizes in the $P4/nmm$ tetragonal space group while a small variation in concentration to FeSe$_{1.06}$ induces a phase change and the material exhibits a hexagonal structure \cite{Schuster1979}.  Furthermore, superconductivity in Fe$_{1.01}$Se with T$_C$ $\sim$ 8 K is destroyed by increasing the amount of excess Fe and no superconductivity is observed down to 0.6 K in samples of Fe$_{1.03}$Se \cite{McQueen2009}.  The phase of interest with respect to superconductivity is the $\alpha$ phase (curiously this phase is occasionally referred to in the literature as the $\beta$ phase).  Magnetic ordering is observed in samples close to the Te endpoint member of the Fe$_{1+y}$Te$_{1-x}$Se$_x$ family. Structurally, Fe$_{1+y}$Te exhibits the PbO crystal structure with a space group of $P4/nmm$ at room temperature for values of $y$ ranging from 0.068 to 0.14 \cite{Bao2009,Li2009g}.  At low temperatures, a first-order structural transition is observed (transition temperature $\sim$65 K \cite{Fang2008b}) and the low temperature space group is the monoclinic $P2_{1}/m$ for samples of Fe$_{1.076}$Te \cite{Bao2009} and Fe$_{1.068}$Te \cite{Li2009g}.  On the other hand, a small change in $x$ to Fe$_{1.141}$Te changes the low temperature unit cell to orthorhombic with the $Pmmn$ space group \cite{Bao2009} again providing evidence for sensitivity to stoichiometry.  In both the orthorhombic and monoclinic unit cells, there is no cell doubling or cell rotation when compared to the tetragonal cell and, hence, the Miller indices of Bragg reflections are the same for all unit cells \cite{Bao2009}.

The magnetic structure and low temperature monoclinic distortion of Fe$_{1.125}$Te was first determined 35 years ago \cite{Fruchart1975}.  The magnetic structure was characterized by an ordering wave vector of (1/2 0 1/2) with a rather large ordered magnetic moment of 2.07 $\mu$B with components of magnetic moment along all crystallographic axes \cite{Fruchart1975}.  Recent neutron diffraction measurements indicate ordering with the same commensurate (1/2 0 1/2) wavevector in the case of Fe$_{1.075}$Te \cite{Bao2009} and Fe$_{1.068}$Te \cite{Li2009g} where the low temperature structure is monoclinic.  The data are consistent with a collinear spin structure \cite{Fruchart1975,Bao2009,Li2009g}.  The more recent studies indicated an ordered moment of 2.03 $\mu$B for $x$=0.075 \cite{Bao2009} and 2.25 $\mu$B for $x$=0.068 \cite{Li2009g} with the majority of the moment along the crystallographic $b$-axis \cite{Bao2009,Li2009g} (Note that the moment direction is different than the earlier magnetic structure determination \cite{Fruchart1975}).  Interestingly, the samples which exhibit a low temperature orthorhombic structure are found to order magnetically with an incommensurate wavevector of ($\pm\delta$,0,1/2) with $\delta\approx$0.38 \cite{Bao2009}.  The addition of Se causes the long-range magnetically ordered state to evolve into short range order centered at incommensurate wavevectors (0.5-$\delta$ 0 0.5) \cite{Bao2009,Wen2009b}.  Single crystal studies of two samples, Fe$_{1.07}$Te$_{0.75}$Se$_{0.25}$ and FeTe$_{0.7}$Se$_{0.3}$ indicate that less excess Fe and more Se makes the incommensuration smaller, such that the scattering is closer to (0.5 0 0.5), and the intensity weaker as well \cite{Wen2009b}.  Interestingly, peaks are only observed on one side of the 2D (0.5 0) wavevector (i.e. the wavevector is (0.5-$\delta$,0,0.5) and not (0.5$\pm\delta$,0,0.5)) although the observed incommensuration is reproduced with a model of exponentially decaying correlations where the characteristic length scale of the exponential is different for ferromagnetic and antiferromagnetic correlations \cite{Wen2009b}.

The wavevector observed in the 11 materials is different than that observed in the 1111 or 122 materials with an in plane wavevector of (1/2 0)$_T$ as opposed to (1/2 1/2)$_T$ seen in both the 1111 and 122 compounds.  Interestingly, this happens despite calculated Fermi surfaces that are very similar suggesting that the 11 materials should be susceptible to a nesting instability with the same (1/2 1/2)$_T$ nesting wavevector \cite{Subedi2008b}.  Calculations indicate that excess Fe in these materials is magnetic \cite{Zhang2009j}, consistent with the conclusions of neutron structure refinements \cite{Bao2009}.  However, these calculations indicate that the lowest energy ground state is still the (1/2 1/2)$_T$ state \cite{Zhang2009j} even in the presence of excess Fe.  It was further proposed that the excess Fe could cause a mismatch in size between the $\Gamma$ and M parts of the Fermi surface which could lead to incommensurate magnetism \cite{Zhang2009}.  Experimentally, ARPES measurements confirmed the calculated Fermi surface and further showed that the $\Gamma$ and M parts of the Fermi surface are closely matched \cite{Xia2009a}.  In addition, despite a weak feature at the X point, these measurements seem to rule out Fermi surface nesting along the observed (1/2 0)$_T$ wavevector.  As Fermi surface nesting seems very unlikely in this case, theoretical approaches to understand the ordering have focussed on local moment models.  First-principles calculations have predicted bicollinear order due to the presence of near neighbor, next-near neighbor and next-next-near neighbor interactions \cite{Ma2009}.  This model was expanded upon and the phase space of the various exchange constants was explored indicating not only ordering with the commensurate (1/2 0)$_T$ wavevector but also incommensurate ordering \cite{Fang2009} as observed experimentally with higher amounts of excess Fe \cite{Bao2009}.  It was also suggested that interaction of the FeTe spins with the magnetic excess Fe moments could modify the superexchange interactions leading to the incommensuration observed experimentally \cite{Fang2009}.  One alternate model, described previously, results in local moments but with magnetic interactions which are long-ranged itinerant interactions and not superexchange \cite{Johannes2009}.  These DFT calculations, when applied to FeTe, indicate that the (1/2 0)$_T$ ordering and (1/2 1/2)$_T$ ordering were energetically very similar as opposed to the 122 case where the (1/2 1/2)$_T$ ground state was clearly favored \cite{Johannes2009}.

\section{Spin Dynamics}

A comprehensive understanding of the relationship between magnetism and superconductivity in the Fe-based superconductors ultimately requires understanding not only the nearby magnetic ground states, but the full magnetic excitation spectrum.  Indeed, once the close proximity of a magnetic state in the phase diagram was discovered theories postulating a magnetic pairing mechanism were quickly put forth (\textit{e.g.} see \cite{Singh2009d,Mazin2009a,Chubukov2009,Kuroki2009}).  Concurrently, the importance of spin fluctuations was established from an experimental point of view where early studies of the spin dynamics \cite{Ahilan2008a,Nakai2008,Christianson2008,Lumsden2009,Chi2009} revealed the presence of strong magnetic fluctuations in the superconducting  region of the phase diagram. Thus, a key step in confirming or rejecting theories of superconductivity relying on a magnetic pairing mechanism requires the elucidation of the magnetic excitation spectrum and thereby an effective Hamiltonian.

In this section the spin dynamics are discussed as revealed primarily by Inelastic Neutron Scattering (INS) and in some cases supplemented by Nuclear Magnetic Resonance (NMR) measurements.  Note that that the two probes are complementary: The NMR energy window is substantially less than INS experiments where energy resolutions are not typically better than 0.1 meV.  On the other hand, INS experiments can probe excitations to as high as 300 meV and beyond.  In section 4.1 the spin dynamics found in the parent compounds are discussed.  Section 4.2 discusses the evolution of the spin dynamics with chemical doping, pressure, and magnetic field.  Finally, section 4.3 is reserved for a comprehensive discussion of the magnetic resonance at various chemical compositions and magnetic fields.

\subsection{Spin Dynamics in the Parent Compounds}

\begin{figure}
\centering
\includegraphics[width=.8\columnwidth]{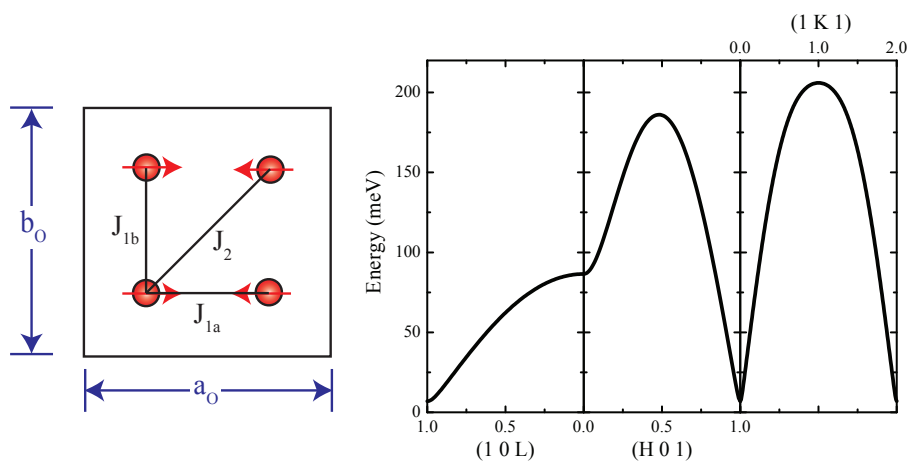}
\caption{\label{dispersion} (a) Exchange constants for the Heisenberg model given in the text.  Note that $J_c$ (not shown) is the nearest neighbor interaction to the Fe ion directly above. (b) Dispersion of CaFe$_2$As$_2$ calculated with the parameters from ref. \cite{Zhao2009}. Note the change in index going from the $(H 0 1)_O$ direction to the $(1 K 1)_O$ direction }
\end{figure}

The present understanding of the spin dynamics in the parent compounds is still evolving. However, many characteristics of the spin excitations are already well established.  Below T$_N$, the spin excitations are characterized by gapped steeply dispersing anisotropic three-dimensional (3D) spin waves extending to high energies.  Above T$_N$ the spin excitations appear to lose much of their 3D character with correlations along the c-axis considerably weakened while the in-plane correlations remain strong.  Because of the large energy scale of both the spin excitations and the gap values most of the current understanding of the spin excitations is derived from analysis of INS data.  Many INS experiments have been performed at energies small compared to the zone boundary energy. In this limit, what is experimentally observed are spin excitations which can be characterized by an antiferromagnetic zone center energy gap and spin wave velocities most commonly along two directions as the majority of experiments performed to date have used a triple-axis spectrometer where measurements are performed in a fixed plane. In order to attempt to extract interaction energies from these velocities, a local moment Heisenberg Hamiltonian has often been adopted.  As the nature of the magnetic interactions (i.e. local vs. itinerant) is still the topic of considerable debate, such a mapping should be taken \emph{cum grano salis}.
For the 122 materials, the Hamiltonian most widely used is \cite{Ewings2008,Yao2008}:
\begin{equation}\label{HHam}
H=\sum_{<jk>}(J_{jk}S_j \cdot S_k)+\sum_j \{D(S_z^{2})_j\}
\end{equation}
Where $J_{jk}$ are exchange constants and $D$ represents a single ion anisotropy term, included to describe the observed energy gap.  The magnetic excitations can be described using three in-plane exchange constants; the near-neighbor interactions $J_{1a}$ and $J_{1b}$ (different due to the orthorhombic distortion) and the next-near-neighbor interaction $J_2$ as shown in Fig. \ref{dispersion}.  There is also a single c-axis nearest neighbor exchange constant $J_c$.  This Hamiltonian results in the following spin wave dispersion
\begin{equation}
\omega_q=\sqrt{A_{q}^{2}-B_{q}^{2}}
\end{equation}
where
\begin{eqnarray}
  A_q &=& 2S\left(J_{1b} (\cos \pi K - 1) + J_{1a} + 2J_2 + J_{1c} + D\right) \\
 \nonumber  B_q &=& 2S\left(J_{1a} \cos \pi H + 2J_2 \cos \pi H \cos \pi K + J_{1c} \cos \pi L\right).
\end{eqnarray}
In the above expressions, $H,K,L$ are the reciprocal space coordinates corresponding to the orthorhombic unit cell.  The majority of the triple-axis experiments have been performed in the ($H$0$L$)$_O$ scattering plane.  In this plane, the spin wave velocities \cite{McQueeney2008} and energy gap are:
\begin{eqnarray}
  v_\parallel &=& aS\left(J_{1a}+2J_2\right)\sqrt{1+J_{1c}/(J_{1a}+2J_2)} \\
\nonumber  v_\perp &=& cSJ_{1c} \sqrt{1+(J_{1a}+2J_2)/J_{1c}}\\
\nonumber  \Delta &=& \sqrt{D\left(D+2(J_{1a}+2J_2+J_{1c}) \right)}
\end{eqnarray}
Note that these observables are insensitive to the value of $J_{1b}$ and only depend on the in-plane interactions as $J_{1a}$-2$J_2$.  Determination of $J_{1b}$ or the unique determination of $J_{1a}$ and $J_2$ requires measurements not restricted to a plane and extending to higher energies as can be obtained from time-of-flight data \cite{Diallo2009,Zhao2009}.

For completeness, we note that INS experiments measure $S(\textbf{Q},\omega)$ which is related to the imaginary part of the dynamic susceptibility, $\chi''(\textbf{Q},\omega)$, by
\begin{equation}
S(\textbf{Q},\omega)=\frac{\chi''(\textbf{Q},\omega)}{\pi(1-e^{-\hbar\omega/k_BT})}
\end{equation}

\begin{figure}
\centering\includegraphics[width=.6\columnwidth]{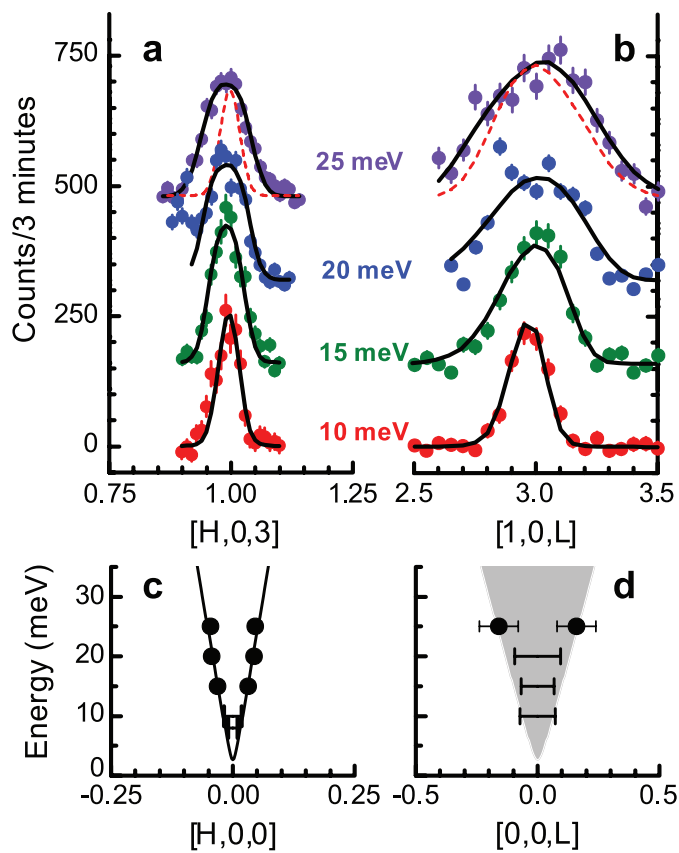}
\caption{\label{McQCaFe2As2} Representative INS scans on CaFe$_2$As$_2$ in the long-wavelength limit (Reprinted with permission from \cite{McQueeney2008}, copyright 2008 the American Physical Society).  Constant energy scans are shown along (a) (H 0 3)$_O$ and (b) (1 0 L) $_O$.  The resulting dispersion is shown in (c) and (d) showing the 3D anisotropic spin waves.}
\end{figure}

We now turn to a discussion of the spin waves in the magnetically ordered state in the 122 family of materials.  The spin excitation spectrum has been measured using INS on samples of BaFe$_2$As$_2$ \cite{Ewings2008,Matan2009}, CaFe$_2$As$_2$ \cite{McQueeney2008,Diallo2009,Zhao2009,Diallo2010}, and SrFe$_2$As$_2$ \cite{Zhao2008c}.  Fig. \ref{McQCaFe2As2} shows representative scans taken from INS measurements of the spin excitations in an array of CaFe$_2$As$_2$ single crystals \cite{McQueeney2008}.  Visual inspection of this data allows for the qualitative understanding of some generic properties of the 122 materials.  In particular, significant dispersion is visible along both the $H$ and $L$ directions.  Hence, the interactions can be characterized as anisotropic 3D interactions, in contrast to the 2D magnetic interactions found in the cuprates \cite{Kastner1998}.  Moreover, it is clear that the spin excitations are steeply dispersive, with an early study by Ewings, \textit{et al.} estimating a bandwidth of 170 meV\cite{Ewings2008} suggesting strong magnetic interactions, as in the cuprates.

Table \ref{tablespinwave} presents a summary of the coupling constants S($J_{1a}+2J_2$), S$J_c$, energy gap, and spin wave velocities extracted from INS on 122 parent compounds (together with the underdoped sample BaFe$_{1.92}$Co$_{0.08}$As$_2$ \cite{Christianson2009}).  The ratio of spin wave velocities (v$_\parallel$/v$_\perp$) is a measure of the anisotropy of the magnetic excitations with an infinite ratio observed for purely in-plane (2D) excitations and unity for isotropic 3D interactions.  As can be seen in table \ref{tablespinwave} the spin wave velocity ratio varies from 1.5 to 5.  Anisotropy in the low energy spin fluctuations is also concluded from NMR data\cite{Kitagawa2008,Kitagawa2009} with BaFe$_2$As$_2$ being more anisotropic than SrFe$_2$As$_2$, a trend that is reflected in Table \ref{tablespinwave}.  Interestingly, the spin wave velocity ratio in BaFe$_2$As$_2$ in notably higher than the Ca and Sr parent compounds indicating more 2D interactions in this material and it is also the Ba-122 compounds that exhibit the highest superconducting transition temperatures for the same dopant (see Table \ref{tctable}).  This may suggest that 2D interactions are favorable for superconductivity.

\begin{table}
\small{
\begin{tabular}{|c|c|c|c|c|c|c|}
  \hline
  Material & S(J$_{1a}$+2J$_2$) & S(J$_{c}$) & Energy Gap  & v$_{\parallel}$ & v$_{\perp}$ & v$_{\parallel}$/v$_{\perp}$\\
  \hline
  CaFe$_2$As$_2$ \cite{McQueeney2008}& 73 $\pm$ 14 & 6.7 $\pm$ 3 & 6.9 $\pm$ 0.2 & 420 $\pm$ 70 & 270 $\pm$ 100 &  1.56\\
  CaFe$_2$As$_2$ \cite{Zhao2009} & 87.7 $\pm$ 12 & 5.3 $\pm$ 1.3 & -  & 498 $\pm$ 70 & 259 $\pm$ 116 & 1.92\\
  CaFe$_2$As$_2$ \cite{Diallo2009} & 91.5 $\pm$ 1.2 & 4.5 $\pm$ 0.1 & - & 516 $\pm$ 7 & 243 $\pm$ 7 &  2.12\\
  SrFe$_2$As$_2$ \cite{Zhao2008c} & 100 $\pm$ 20 & 5 $\pm$ 1 & 6.5 & 560 $\pm$ 100 & 280 $\pm$ 56 &  2 \\
  BaFe$_2$As$_2$ \cite{Matan2009} & 50 $\pm$ 27 & 0.38 $\pm$ 0.15 & 9.8 $\pm$ 0.4 & 280 $\pm$ 150 & 57 $\pm$ 7 &  4.91\\
  BaFe$_{1.92}$Co$_{0.08}$As$_2$ \cite{Christianson2009} & 32 $\pm$ 1 & 0.34 $\pm$ 0.01 & 8.1 $\pm$ 0.2 & 180 $\pm$ 12 & 43 $\pm$ 2 & 4.19\\
  \hline
\end{tabular}
}
\caption{Exchange constants extracted from inelastic neutron scattering on 122 single crystals using an effective Heisengberg Hamiltonian.  All energies are reported in units of meV and spin wave velocities in units of meV \AA.  The value of the energy gap is lineshape dependent and thus care should be taken when making comparisons between studies.}
\label{tablespinwave}
\end{table}

Another prominent feature of the spin wave data in the parent compounds is the presence of a gap in the excitation spectrum (see Table \ref{tablespinwave} for the gap extracted from INS measurements). The opening of a gap in the spin excitation spectrum is also reflected in the temperature dependence of the NMR data (\textit{e.g.}\cite{Fukazawa}). The gap largely disappears above T$_N$ \cite{Zhao2008c,Diallo2010} while well below T$_N$ the gap appears to be rather broad \cite{Ewings2008,Matan2009}.  Ewings \textit{et al.} put forth idea that multiple gaps may exist that experimental resolution is insufficient to resolve.  While a single ion anisotropy is included in the effective Heisenberg Hamiltonian used to extract the gap, this is really \textit{ad hoc} and, thus, an unresolved question is the nature and origin of the gap in the spin wave spectrum.

As mentioned above, measurements performed in the long-wavelength limit are insensitive to $J_{1b}$ and, consequently, the in-plane anisotropy cannot be determined.  This situation can be remedied by measurements which extend to the zone boundary.
Such measurements using time-of-flight INS were performed on single crystals of CaFe$_2$As$_2$ resulting in effective nearest neighbor exchange interactions of $SJ_{1a}\sim50(10)$meV, $SJ_{1b}\sim -6(5)$meV and a rather large next nearest exchange interaction of $SJ_2$ of 19(3) meV \cite{Zhao2009}.  The dispersion (Eqs. 2 and 3) calculated from these parameters is shown in Fig. \ref{dispersion}.
The observation of antiferromagnetic $J_{1a}$ and $J_2$ and ferromagnetic $J_{1b}$ shows that, unlike theoretical predictions \cite{Yildirim2008,Ma2009a,Si2008}, the magnetic interactions are not frustrated.  However, the relatively large size of $J_2$ is consistent with theory \cite{Yildirim2008,Ma2009a,Si2008} and the measured ratio of $J_{1a}$/2$J_2$ is $\sim$ 1.3(3).
The anisotropy in the nearest neighbor exchange constants is striking for a system that is tetragonal at room temperature.
Similar anisotropy was predicted from a combination of DFT calculations and linear response theory suggesting rather short range interactions and exchange constants close to experimental observation (for instance, the calculated $J_{1a}$/2$J_2$ ratio is $\sim$ 1 for various FeAs parent compounds) \cite{Han2009b}.  An alternative model was proposed which involves breaking the in-plane rotational symmetry via orbital ordering suggesting that orbital degrees of freedom may play an important role in these materials \cite{Lee2009f}.  Further experimental work, particulary with full control over all three reciprocal space coordinates is crucial in this area.

The experimental situation in the parent compound in other families of materials is much less settled, primarily do the the lack of large single crystals.  To our knowledge, there have only been two investigations of the spin excitations in the 1111 family of materials.  Despite the limitations of polycrystalline samples and the resulting spherical averaging of the spin excitation spectrum, the studies showed a spectrum qualitatively similar to that observed in the 122 parent compounds.  Measurements on LaFeAsO \cite{Ishikado2009} revealed a column of spin wave excitations emanating from a wave vector consistent with $(1 0 L)_O$.   Well above the magnetic ordering temperature the peak shape exhibits an asymmetry in Q interpreted as being consistent with two-dimensional (2D) spin fluctuations.  Given this as well as the fact that the excitations are steeply dispersing, these measurements are consistent with strong in-plane exchange interactions between Fe atoms \cite{Ishikado2009}.
Further work on the Fe-spin excitations in the other 1111 has been hampered by the overlap with crystal field excitations of the rare earth ions.  The crystal field excitations in the 1111 materials have been explored for the case of CeFeAsO$_{1-x}$F$_x$ \cite{Chi2008}.  In the disordered state, the observed crystal field excitations are consistent with the expected system of three doublets for Ce$^{3+}$ in a tetragonal environment \cite{Chi2008}.  Below the Fe magnetic ordering temperature, the degeneracy of the doublets is lifted due to the internal field of the magnetically ordered Fe-sublattice.  Finally, in the 11 family, limited data exists with measurements on a polycrystalline sample of FeTe$_{0.92}$\cite{Iikubo2009} indicating spin excitations at a Q consistent with the magnetic ordering (0.5 0 0.5) wave vector \cite{Bao2009}.

In principal, the study of the spin excitations can provide information concerning whether or not the spin degrees of freedom are derived from localized or itinerant electrons.  In an itinerant antiferromagnet, there should be significant damping due to decay of the spin waves into electron-hole pairs.
Measurements on CaFe$_2$As$_2$ \cite{McQueeney2008} and BaFe$_2$As$_2$ \cite{Matan2009} suggested that considerable damping is present in the spin wave spectrum as expected for an itinerant system.  On the other hand, measurements of the full excitation spectrum in CaFe$_2$As$_2$ are well described by a Heisenberg Hamiltonian \cite{Zhao2009} with some damping ($\Gamma$ $\sim$ 0.15 E) which they attribute to predominately local moment physics.  Interestingly, despite the contradictory conclusions with respect to itineracy, the magnitude of damping extracted by McQueeney \emph{et al.} \cite{McQueeney2008} and Zhao \emph{et al.} \cite{Zhao2009} both on CaFe$_2$As$_2$ are largely consistent with one another.  Ultimately there may well be aspects of both localized and itinerant magnetism as predicted from DFT calculations where it was suggested that the moments were local but the interactions were itinerant in nature \cite{Johannes2009}.

\subsection{Evolution of the Spin Excitations}

We now turn to the evolution of the spin excitations as a function of a tuning parameter.   Much of the experimental effort so far has been directed towards understanding the spin resonance which appears in the spin excitation spectrum below T$_C$ which will be discussed in detail in the following section.  Here we concentrate on the more general evolution of the spin excitations.  In fact, the normal state spin excitations, assuming a magnetically mediated pairing mechanism, must contain the necessary spectral weight to facilitate pairing.  The most common tuning parameter to date has been chemical doping, though applied magnetic fields and pressures have also been used.  In this section we will largely limit discussion to three main classes of materials (Ba(Fe,Co,Ni)$_2$As$_2$, LaFeAs(O,F), and Fe(Te,Se)) as they exemplify the characteristic behavior of the Fe-based materials.

\begin{figure}
\centering\includegraphics[width=.9\columnwidth]{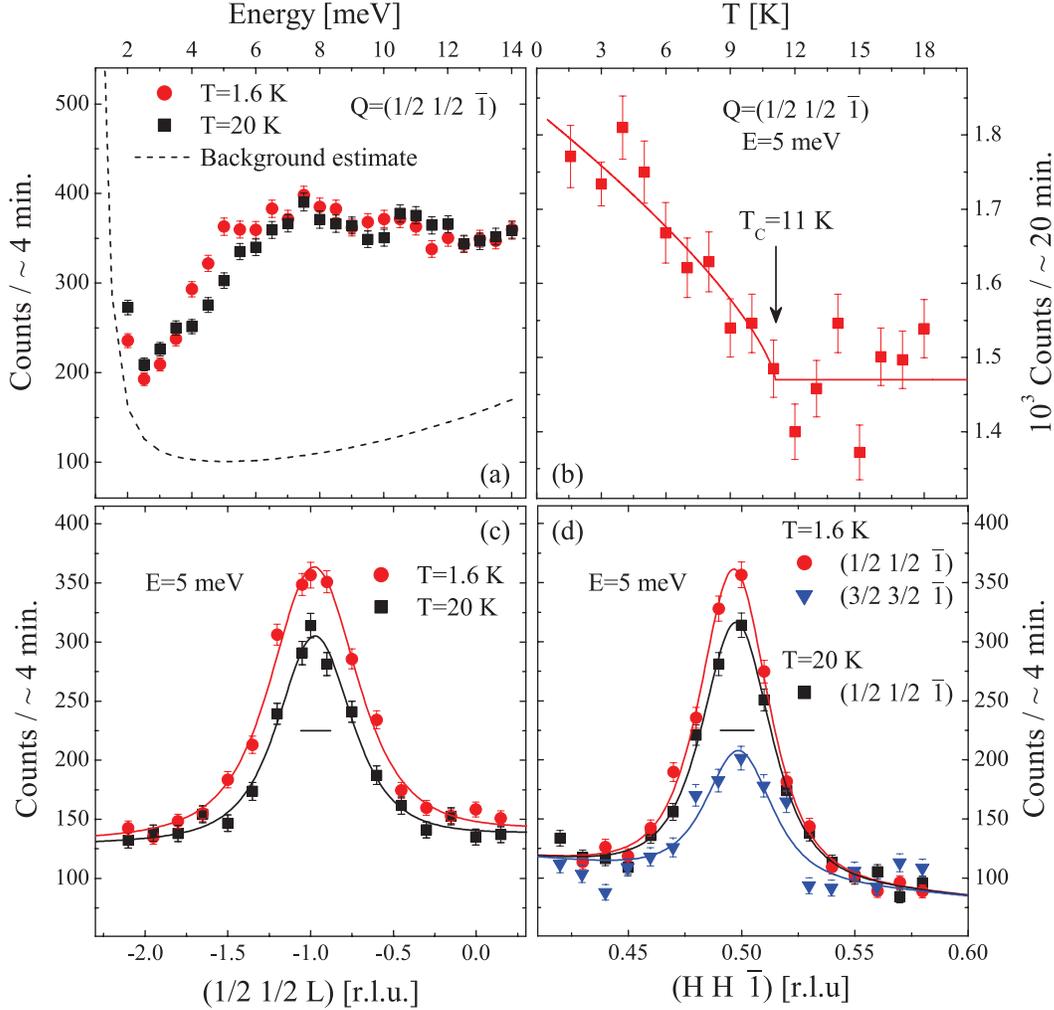}
\caption{\label{underdoped_res} INS data for underdoped BaFe$_{1.92}$Co$_{0.08}$As$_2$ (Reprinted with permission from \cite{Christianson2009}, copyright 2009 the American Physical Society). (a) Constant-Q scans, Q=(1/2 1/2 -1)$_T$, for temperatures above and below T$_C$ together with the estimated background.  (b)The temperature dependence of the inelastic intensity at (1/2 1/2 -1)$_T$ and E=5 meV.  The solid line is a power law fit yielding T$_C$=11(1) K.
L-dependence (c) and H-dependence (d) of the inelastic intensity near Q=(1/2 1/2 -1)$_T$ and E=5 meV at 1.6 and 20 K.
The solid lines are guides to the eye and horizontal bars represent instrumental resolution.
In (d) a scan around (3/2 3/2 -1)$_T$ is included to emphasize the magnetic origin of the scattering.}
\end{figure}

The spin dynamics in the Ba(Fe$_{1-x}$Co$_x$)$_2$As$_2$ has been studied by several groups\cite{Lumsden2009,Pratt2009,Christianson2009,Inosov2010,Matan2010, Lester2010}.  In the underdoped region,
the spin excitations of a single crystal sample of magnetically ordered (T$_N$=58 K) and superconducting (T$_C$=11 K) BaFe$_{1.92}$Co$_{0.08}$As$_2$ were studied in the long-wavelength limit \cite{Christianson2009}.  In accord with the results in BaFe$_2$As$_2$ the excitations are
peaked in both H and L so that it is clear that the magnetic interactions, though anisotropic, are 3D (see Fig. \ref{underdoped_res}).  The anisotropy of the magnetic interactions as characterized by the ratio
of the $v_{\|}/v_{\bot}$ is about 4.2 consistent with the value found in the BaFe$_2$As$_2$ \cite{Matan2009}.  The low temperature spin waves are characterized by a gap ($\Delta$ $\sim$ 8 meV), comparable to that found in the parent compound BaFe$_2$As$_2$ \cite{Matan2009}.  Note, however, that care must be taken when making such comparisons as the gap energy is lineshape dependent.

Optimally doped BaFe$_2$As$_2$ has been studied for the case of Co and Ni doping.  As discussed earlier, in the optimally doped region of the phase diagram long range magnetic order has been suppressed (see fig. \ref{phasediagram}).  Initial studies of BaFe$_{1.84}$Co$_{0.16}$As$_2$ (T$_C$=22 K) with INS showed a striking reduction of the spin correlations along the c-axis as shown by a visual inspection of in-plane and c-axis scans \cite{Lumsden2009}.
The ratio of the in-plane to c-axis bandwidths allows a more quantitative determination of the degree of two-dimensionality.  To that end, data on BaFe$_{1.84}$Co$_{0.16}$As$_2$ was analyzed with a dispersion composed of a gap with in-plane and c-axis coupling constants.  From this model an estimate of the ratio of in-plane to c-axis bandwidths of 117 (with a lower limit of 40) was determined.  This ratio is much larger than that of the parent compounds (see Fig. \ref{dispersion} for CaFe$_2$As$_2$ and the spin wave velocity ratio of Table \ref{tablespinwave}). This confirms a substantial reduction in c-axis correlations relative to those in the tetragonal plane.
Thus, the dimensionality of the magnetic interactions has shifted from anisotropic 3D in the parent compounds towards 2D.  Although INS measurements on optimally doped Ba$_{1.9}$Ni$_{0.1}$As$_2$ (T$_C$=20 K) are primarily focussed on elucidating the behavior of the spin resonance, they also observe a reduction in the magnetic correlations along the c-axis \cite{Chi2009}.  This reduction is apparently not as large as for the case of Ba$_{1.84}$Co$_{0.16}$As$_2$, but a more quantitative analysis of both the in-plane and c-axis dispersion is required before a firm comparison can be made. The spin excitations of Ba$_{1.84}$Co$_{0.16}$As$_2$ with T$_C$=25 were followed to high temperatures and up to 35 meV \cite{Inosov2010} and the resulting spectrum analyzed with a model for a nearly antiferromagnetic Fermi liquid \cite{Moriya1985}
The data were normalized to be in absolute units and that they were able to extract a total spectral weight in the normal state spin fluctuations of $0.17 \mu_B^2/f.u.$ up to 35 meV \cite{Inosov2010}.  This value is comparable to that found for underdoped YBa$_2$Cu$_3$O$_{6+x}$ \cite{Fong2000}.

There have been somewhat fewer studies of the spin dynamics in the 122 materials in the overdoped region.  NMR investigations have probed the evolution of the spin excitations into the overdoped region in Ba(Fe$_{1-x}$Co$_x$)$_2$As$_2$ \cite{Ning2010}.  These measurements find that as T$_C$ is suppressed by Co-doping so are the low energy spin fluctuations.   Higher energy spin excitations in the Ba(Fe$_{1-x}$Co$_x$)$_2$As$_2$ series have been studied with INS into the overdoped region \cite{Matan2010}.  In this region two concentrations, $x$=0.14 (T$_C=7$ K) and $x$=0.24 (T$_C=0$) are studied.  For $x$=0.14 intensity is observed at (1 0 1)$_O$ that appears to be consistent with gapped spin excitations,  at 10, 30 and 100 K, which is substantially different from the more lightly doped samples where a gap in the spin excitations does not appear until the onset of long range magnetic order or superconductivity.  The observation of gapped spin excitations may also explain the disappearance of spin fluctuations at $x>$0.15 seen with NMR \cite{Ning2010} as such a gap would effectively redistribute the spectral weight outside the NMR measurement window.  Matan, \textit{et al.} also examined a sample where superconductivity has been fully suppressed by doping and were unable to observe spin fluctuations \cite{Matan2010}.  They argue that this observation is further evidence that the magnetic excitations are due to nesting between the hole and electron Fermi surfaces as the hole pocket should have effectively disappeared at this concentration \cite{Brouet2009}.
On the other hand, recent NMR studies of hole doped Ba$_{1-x}$K$_x$Fe$_2$As$_2$ \cite{Zhang2010} are consistent with the development of a new type of spin fluctuation developing near KFe$_2$As$_2$ perhaps suggesting a different pairing symmetry on the overdoped side of the phase diagram\cite{Zhang2010} .

The available studies of the LaFeAsO$_{1-x}$F$_x$ series are also consistent with the disappearance of spin fluctuations as the hole pocket is filled by electron doping. The low energy spin fluctuations in polycrystalline samples of LaFeAsO$_{1-x}$F$_x$ were explored with INS \cite{Wakimoto2010}.  They find similar spin excitations to those found in LaFeAsO\cite{Ishikado2009}.  However, in the overdoped region, the spin fluctuations are observed to have nearly vanished in accord with a close relationship between spin fluctuations and superconductivity.  Thus, in at least two cases (Ba122 and La1111) when antiferromagnetically ordered parent compounds are overdoped by electron doping the spin fluctuations vanish as the hole pocket vanishes.

CaFe$_2$As$_2$ provides an interesting case to study the evolution of the spin excitations as a function of pressure.  Initially, it was thought that CaFe$_2$As$_2$ exhibited pressure induced superconductivity at relatively low pressures \cite{Torikachvili2008,Park2008}.  More recent studies under hydrostatic pressure do not support bulk superconductivity in the high pressure state \cite{Yu2009b}.  Given this, the neutron scattering studies as a function of pressure \cite{Pratt2009a} show that in the pressure region of the collapsed tetragonal phase where long range magnetic order has been suppresed \cite{Goldman2009} the spin fluctuations are also absent in contrast to chemical doping in the 122 materials where strong spin fluctuations are observed despite the suppression of long range magnetic order.   In this case, superconductivity is not supported in the absence of spin fluctuations.

The spin excitations in FeTe$_{1-x}$Se$_x$ have been studied by a number of groups \cite{Mook1,Iikubo2009,Qiu2009,Lumsden2010,Argyriou2010,Mook2,Lee2010}.  The majority of the investigations of the spin excitations in FeTe$_{1-x}$Se$_x$ have focussed on investigation of the magnetic resonance feature which will be discussed in greater detail in the next section.  An interesting early development in the study of the spin excitations was that, in contrast to the parent compound where the magnetic order occurs near (0.5 0 0.5)$_T$, the spin fluctuations in superconducting samples were found to be near (0.5 0.5 0)$_T$ \cite{Mook1,Iikubo2009,Qiu2009} (see fig. \ref{magstructure}).  Thus, the spin correlations in the plane were at a similar wave vector as found in the other Fe-based materials perhaps hinting at the fact that spin fluctuations of a certain type are needed for superconductivity in the Fe-based materials.

\begin{figure}
\centering\includegraphics[width=.9\columnwidth]{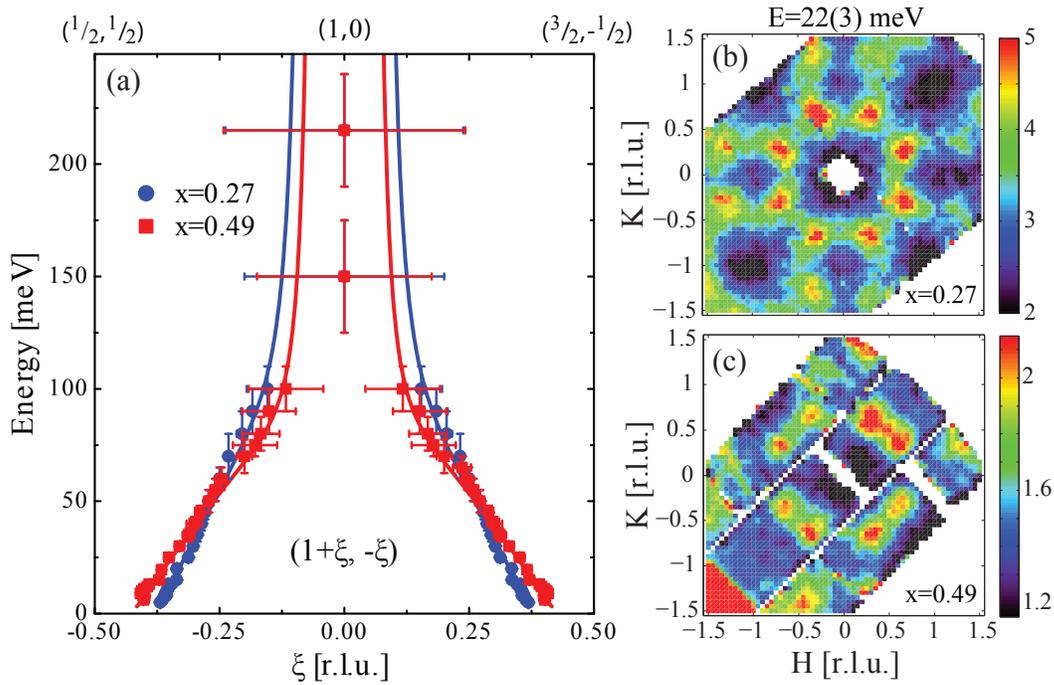}
\caption{\label{FeSeTe} Spin Excitations in FeTe$_{1-x}$Se$_x$ (Reprinted by permission from Macmillan Publishers Ltd: Nature Physics \cite{Lumsden2010}, copyright 2010) .  (a)Dispersion of the spin excitations for x=0.27 (0.49) blue (red). (b) and (c) are constant energy slices of the scattering in the (H K 0)$_T$ plane for x=0.27 (b) and for x=(0.49) (c)}
\end{figure}

Lumsden, \textit{et al.} followed the spin excitations up to 250 meV in single crystal samples \cite{Lumsden2010} (see fig. \ref{FeSeTe}).  A superconducting sample with a T$_C$ of 14 K (FeTe$_{0.51}$Se$_{0.49}$) as well as a sample which did not exhibit bulk superconductivity (Fe$_{1.04}$Te$_{0.73}$Se$_{0.27}$) were measured with INS enabling comparisons of the spin excitations for the two cases.
Both samples showed spin excitations which were 2D in character consistent with measurements over a more limited energy range in FeTe$_{0.6}$Se$_{0.4}$ \cite{Qiu2009}.  The observed excitations originate from a wavevector near (0.5 0.5 0)$_T$ but the zero energy extrapolation of the dispersion indicates that the excitations are incommensurate forming a low energy quartet of peaks characterized by the wavevectors (1$\pm\xi$,$\pm\xi$,L)$_T$ and (1$\pm\xi$,$\mp\xi$,L)$_T$ \cite{Lumsden2010}.  Thus, the excitations exhibit four-fold symmetry around the (1 0 0)$_T$ wavevector as opposed to (1/2 1/2 L)$_T$, the wavevector of magnetic order in the parent compound.  The observed periodicity of the spin excitations is consistent with the 2D square lattice of Fe atoms indicating that this unit cell contains the necessary information to describe the magnetism. More recent measurements have been performed on a sample of FeTe$_{0.6}$Se$_{0.4}$ confirming the incommensurate nature of the spin excitations \cite{Argyriou2010}.    More recently, several observations have appeared that suggest that the symmetry of the spin excitations observed in the FeTe$_{1-x}$Se$_x$ system is common to the 122 materials as well.  In optimally doped Ba(Fe,Co)$_2$As$_2$ two groups have observed a discrete set of spots with an in-plane anisotropy similar to that observed in FeTe$_{1-x}$Se$_x$  rather than a cone as might be expected for a conventional spin wave \cite{Lester2010,Li_HF2010}.  Similar observations appear to hold for the spin excitations in the paramagnetic state of CaFe$_2$As$_2$ \cite{Diallo2010} where a smaller anisotropy of the spin excitations around the (0.5 0.5 L)$_T$  has been observed.

The observation of spin excitations that are four-fold symmetric about the (1 0) (square lattice ($\pi$ $\pi$)) wavevector (see Fig. \ref{FeSeTe}) \cite{Lumsden2010} is intriguing since the excitation spectrum of the cuprates also shows four-fold symmetry about this wavevector.  This indicates strong similarities in the excitation spectrum of Fe- and Cu-based superconductors which, under the assumption of magnetically mediated superconductivity, may suggest a common origin of superconductivity. As the spin excitations in superconducting FeTe$_{0.51}$Se$_{0.49}$ emanate from a position closer to the (0.5 0.5 0)$_T$ than do those from nonsuperconducting Fe$_{1.04}$Te$_{0.73}$Se$_{0.27}$, it is tempting to conclude that spin excitations near (0.5 0.5 0)$_T$ are important for pairing.   Unfortunately, this interpretation is clouded by the presence of the interstitial Fe in Fe$_{1.04}$Te$_{0.73}$Se$_{0.27}$ which may be pair breaking \cite{Zhang2009j}.

\subsection{The Spin Resonance}

The origin and importance of a spin resonance in the INS spectrum is the subject of considerable debate, \textit{e.g.} \cite{Monthoux1994,Fong1995,Batista2001,LeePA2006,Chang2007_res,Norman2007}.  Experimentally, a resonance in the spin excitation spectrum occurring at the onset of T$_C$ was first found in the cuprates \cite{Rossat1991,Mook1993,Fong1999,Dai2000} and subsequently in the heavy fermion materials \cite{Sato2001,Stock2008}.   A similar phenomena was also discovered in INS studies of a polycrystalline sample of Ba$_{1-x}$K$_x$Fe$_2$As$_2$ \cite{Christianson2008}.  This work was subsequently followed by observations of a spin resonance in single crystals of BaFe$_{1.84}$Co$_{0.16}$As$_2$\cite{Lumsden2009} and BaFe$_{1.9}$Ni$_{0.1}$As$_2$\cite{Chi2009}.  Following that numerous observations of a spin resonance in various Fe-based superconductors have been made\cite{Christianson2008,Lumsden2009,Chi2009,Li2009e,Parshall2009,Inosov2010,Mook1,Qiu2009,Argyriou2010,Mook2,Wen2010,Shamoto2010,Christianson2009,Lee2010} (see table \ref{reson}).

The resonance is manifested in the INS spectrum as the appearance of new intensity below T$_C$ localized in both wave vector and energy.  This additional intensity exhibits a temperature dependence which is strongly coupled to the onset of superconductivity (\textit{e.g.} see fig. \ref{underdoped_res}(b)) and the spectral weight appears to arise from a gapping of the spin excitation spectrum.  Theoretically, a spin resonance occurs because of the enhancement of the dynamic susceptibility through a sign change of the superconducting order parameter on different parts of the Fermi surface \cite{Monthoux1994, Fong1995,Chang2007_res,Norman2007,Osborn2009} and consequently the observation of a spin resonance is typically taken as strong evidence for an unconventional pairing symmetry such as \textit{d}-wave or extended \textit{s}-wave ($s$$\pm$).

\begin{figure}
\centering\includegraphics[width=.9\columnwidth]{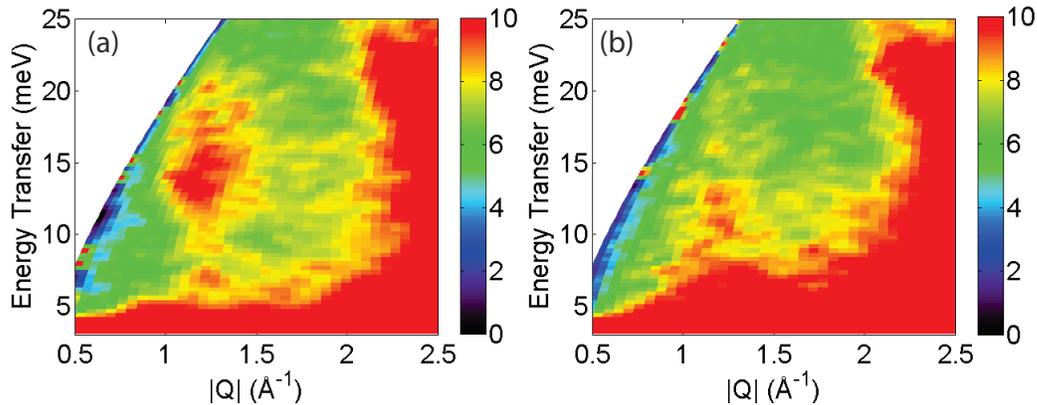}
\caption{\label{polyres} Resonant spin excitation in Ba$_{0.6}$K$_{0.4}$Fe$_2$As$_2$. (Reprinted by permission from Macmillan Publishers Ltd: Nature \cite{Christianson2008}, copyright 2008)  (a) INS data for T$<$T$_C$. (b) INS data for T$>$T$_C$.  The spin resonance is visible as the bright spot in (a).}
\end{figure}

As noted above, a spin resonance in the Fe-based family of superconductors was first observed in polycrystalline samples of Ba$_{0.6}$K$_{0.4}$Fe$_2$As$_2$ \cite{Christianson2008}.  Figure \ref{polyres} shows INS data as a function of Q and energy transfer for temperatures below (a) and above (b) T$_C$.  The resonance is the clearly identifiable spot of intensity below T$_C$.  The resonant intensity appears at a wavevector consistent with (1/2 1/2 0)$_T$ suggesting that the resonance occurs due to dynamic spin correlations in the tetragonal basal plane. In-plane spin correlations are consistent with NMR measurements in optimally doped Ba$_{0.6}$K$_{0.4}$Fe$_2$As$_2$ \cite{Yashima2009}. However, a unique identification of the wave vector is not possible without INS measurements on a single crystal specimen.

The discovery of superconductivity on replacement of Fe with Co in BaFe$_2$As$_2$ \cite{Sefat2008b} allowed significant progress to be made due to the availability of relatively large single crystals.  Shortly thereafter, INS experiments were reported on optimally doped BaFe$_{1.84}$Co$_{0.16}$As$_2$ (T$_C$=22 K) \cite{Lumsden2009} and BaFe$_{1.9}$Ni$_{0.1}$As$_2$ (T$_C$=20 K)\cite{Chi2009}.  These measurements confirmed the first results on Ba$_{0.6}$K$_{0.4}$Fe$_2$As$_2$ and provided several new pieces of information.  The excitations of BaFe$_{1.84}$Co$_{0.16}$As$_2$ were explored with a combination of time-of-flight and triple-axis techniques which located the spin excitations with high confidence at the $(0.5 ~0.5 ~L)_T$ wave vectors \cite{Lumsden2009}.  Triple-axis measurements on BaFe$_{1.9}$Ni$_{0.1}$As$_2$  performed in the (HHL)$_T$ plane are also consistent with this wavevector\cite{Chi2009}.  As discussed above, the $c$-axis spin correlations were found in both cases to be much weaker than those observed in the parent compounds.  The resonance appears to mimic the behavior of the spin excitations in this regard with only a weak dependence on L in BaFe$_{1.84}$Co$_{0.16}$As$_2$\cite{Lumsden2009} and perhaps a somewhat stronger L-dependence in BaFe$_{1.9}$Ni$_{0.1}$As$_2$ \cite{Chi2009,Li2009e}.  The work in both materials showed the appearance of a gap in the spin fluctuation spectrum coincident with the onset of T$_C$ and, thus, the spectral weight in the resonance appeared to be derived from a redistribution of spectral weight from low energy to the resonance position.

The opening of the superconducting gap with temperature and the relationship to superconductivity was more completely explored in BaFe$_{1.9}$Ni$_{0.1}$As$_2$ \cite{Li2009e}.  The authors argue that the opening of a gap in the spin excitations spectrum is temperature dependent and evolves in the same manner as angle resolved photoemission \cite{Terashima2009}.  The idea that the opening of a gap in the spin excitation spectrum mirrors the opening of the superconducting gap was further clarified by INS measurements of an optimally doped sample with composition BaFe$_{1.85}$Co$_{0.15}$As$_2$ (T$_C$=25 K) where the temperature dependence of the resonance energy was found to be the same as that of the superconducting gap \cite{Inosov2010}.

\begin{figure}
\centering\includegraphics[width=.9\columnwidth]{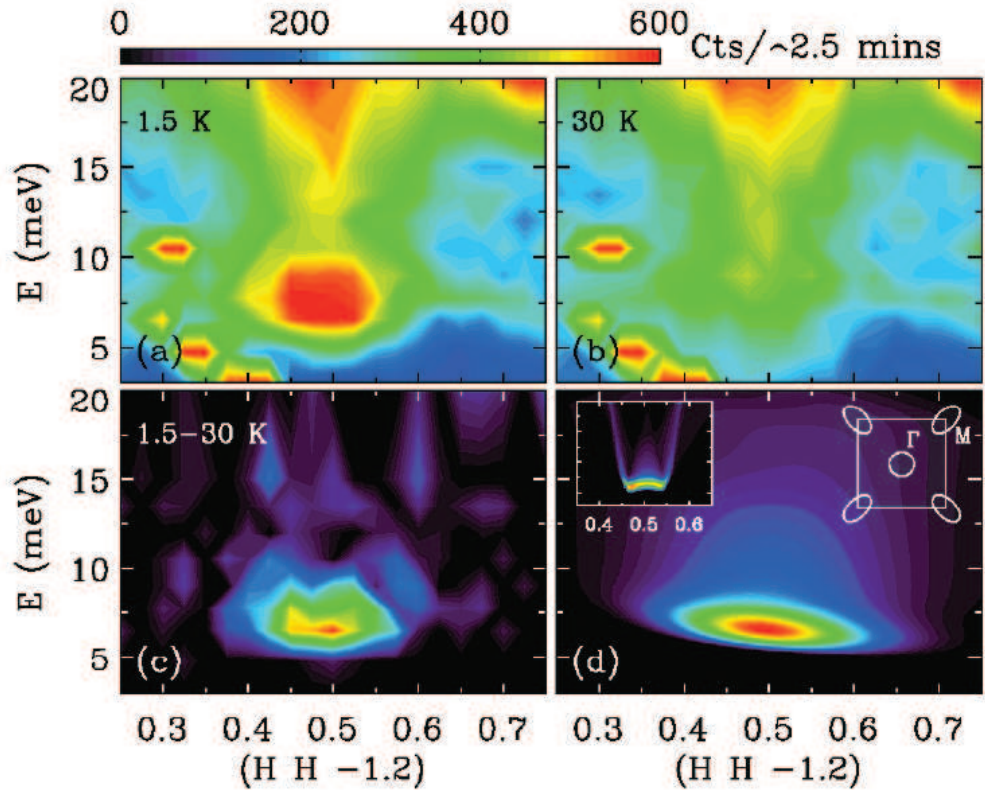}
\caption{\label{FeSeTe_res} Spin resonance in FeTe$_{0.6}$Se$_{0.4}$ (Reprinted with permission from \cite{Qiu2009}, copyright 2009 the American Physical Society).  (a) and (b) shows the spin excitation spectrum as a function of Q and energy at 1.5 and 30 K respectively.  (c) Shows the resonant intensity as determined by the difference between the 1.5 and 30 K spectra.  (d) a theoretical calculation for the resonant intensity as described in \cite{Qiu2009}.}
\end{figure}

A spin resonance has also been observed in several other Fe-based superconductors including: LaFeAsO$_{0.918}$F$_{0.082}$ \cite{Shamoto2010} and several samples in the FeTe$_{1-x}$Se$_x$ series\cite{Mook1,Qiu2009,Argyriou2010,Mook2,Wen2010,Lee2010}.  Measurements on polycrystalline samples of LaFeAsO$_{0.918}$F$_{0.082}$ show the existence of the resonance in the 1111 materials with a resonance energy of 12.9(1) meV corresponding to 5.2 k$_B$T$_C$ \cite{Shamoto2010}. In the FeTe$_{1-x}$Se$_x$ series, the observation of a spin resonance was first reported by Mook \textit{et al.}\cite{Mook1} and Qiu, \textit{et al.} \cite{Qiu2009} (see Fig. \ref{FeSeTe_res}).  Both studies showed a resonance located at the (0.5 0.5 0)$_T$ position in the (HHL) scattering plane; a position in common with that in the 122 materials but not at the position of the ordering wave vector in the parent compound FeTe \cite{Fruchart1975,Bao2009,Li2009g}.  Qiu, \textit{et al.} were able to show that the resonance was 2D in character and hence was due to in-plane spin correlations \cite{Qiu2009}.  As discussed above, the normal state spin fluctuations out of which the resonance evolves are peaked near (1/2 1/2 0)$_T$ but are actually incommensurate in the orthogonal direction \cite{Lumsden2010}. Further studies have subsequently examined the relationship of the location of the spin resonance and normal state incommensurate spin excitations \cite{Mook2,Argyriou2010,Lee2010}.

\begin{table}
\caption{\label{reson} Resonance Energies}
\begin{tabular}{|c|c|c|c|c|c|c|}
\hline
\centering{Material} & T$_C$ (K) & E$_R$ (meV) & E$_R$ (k$_B$T$_C$) & E$_R$/$2\Delta_{max}$ & Ref.\\
\hline
Ba$_{0.6}$K$_{0.4}$Fe$_2$As$_2$& 38  & 14 & 4.3 & 0.58  & \cite{Christianson2008,Ding2008}  \\
LaFeAsO$_{0.918}$F$_{0.082}$ & 29 & 12.9(1) & 5.2 & -  & \cite{Shamoto2010}  \\
BaFe$_{1.85}$Co$_{0.15}$As$_2$ & 25 & 9.5 & 4.4 & 0.79  & \cite{Inosov2010,Terashima2009,Samuely2009,Yin2009} \\
BaFe$_{1.84}$Co$_{0.16}$As$_2$& 22 & 8.6(5) & 4.5 & 0.69  & \cite{Lumsden2009,Yin2009}\\
BaFe$_{1.9}$Ni$_{0.1}$As$_2$& 20  & 7.0(5)-9.1(4) & $\sim$4.7& -  & \cite{Chi2009,Li2009a} \\
FeTe$_{0.5}$Se$_{0.5}$& 14  & 6-7 & $\sim$5.4 & -  & \cite{Mook1,Mook2}\\
FeTe$_{0.5}$Se$_{0.5}$ & 14 & 6-6.5 & $\sim$5.2 & -  & \cite{Lee2010,Wen2010} \\
FeTe$_{0.6}$Se$_{0.4}$& 14 & 6.51(4) & 5.4 & -  & \cite{Qiu2009,Argyriou2010} \\
BaFe$_{1.92}$Co$_{0.08}$As$_2$& 11& 4.5(5) & 4.7 & -  & \cite{Christianson2009}\\
CeCoIn$_5$& 2.3 & 0.60(3) & 3.0 & 0.65  & \cite{Stock2008}\\
\hline
\end{tabular}
\caption{Comparison of spin resonance energies in various Fe-based materials. The references for the experimental determination of the superconducting gap are also given, where available, in the Ref. column.  The heavy Fermion superconductor CeCoIn$_5$ is included for comparison.}
\end{table}

As the origin and importance of the resonance are the subjects of considerable debate, probing the nature of the spin resonance as a function of a various tuning parameters is critical.   To date this has been done with both chemical doping and applied magnetic fields in the Fe-based superconductors. Underdoped samples of BaFe$_{1.92}$Co$_{0.08}$As$_2$ (T$_C$=11 K, T$_N$=58 K) \cite{Christianson2009} (see Fig. \ref{underdoped_res}) and BaFe$_{1.906}$Co$_{0.094}$As$_2$ (T$_C$=17, T$_N$=47 K) \cite{Pratt2009} both show the presence of a spin resonance coexisting with the spin waves of the magnetically ordered state.
Analysis of the spin waves in BaFe$_{1.92}$Co$_{0.08}$As$_2$ are consistent with a gapped spectrum (for temperatures below T$_N$ and above T$_C$) as in the parent compounds \cite{Christianson2008}.  Note that the observation of a gap is complicated by the presence of significant damping.  Below T$_C$, there does not appear to be a suppression of spectral weight at energies below the resonance unless such suppression occurs below $\sim$2 meV.  Thus, unlike the optimally doped materials, the spectral weight for the resonance appears to derive from a source other than the gapping of the spin fluctuations below T$_C$.  Interestingly, the static antiferromagnetic order may prove to be the origin of the spectral weight in the resonance as the magnetic Bragg peaks exhibit suppressed spectral weight below T$_C$\cite{Christianson2009,Pratt2009} (see Fig. \ref{underdopeddiff}).  A similar interplay between magnetism and superconductivity has been seen in heavy Fermion superconductor UPd$_2$Al$_3$\cite{Metoki1998}.

One prominent viewpoint is that the spin resonance is a singlet-triplet excitation.  This hypothesis can be tested by experiments conducted with applied magnetic field which should in principle lift the degeneracy of the triplet excited state \cite{Dai2000,Tranquada2004}. The effect of an applied magnetic field has been studied in BaFe$_{1.9}$Ni$_{0.1}$As$_2$ \cite{Zhao2010} and FeTe$_{0.5}$Se$_{0.5}$ \cite{Wen2010,Qiu2009}.
No change in the resonance was seen in FeTe$_{0.6}$Se$_{0.4}$ in the presence of a 7 T applied magnetic field \cite{Qiu2009} while measurements on FeTe$_{0.5}$Se$_0.5$ showed an intensity change in a 7 T field with no detectable change in resonance energy \cite{Wen2010}.  The change in the resonance intensity is attributed to changes in the superconducting volume \cite{Wen2010}.
The resonance in BaFe$_{1.9}$Ni$_{0.1}$As$_2$ has been measured in magnetic fields up to 14.5 T \cite{Zhao2010}.  Comparing the 14.5 T data to the zero field data, the resonance is reduced in intensity, shifted downward in energy, and slightly broadened.  The experimental data is consistent with the resonance being directly correlated with the superconducting gap \cite{Zhao2010}.  Very recently (just prior to final submission of this article), measurements of the FeSe$_{0.4}$Te$_{0.6}$ in applied fields up to 14 T have been reported \cite{BaoW2010}.  These measurements appear to show a field induced peak splitting and may be consistent with a singlet-triplet excitation.  Certainly, further experimental work in this area is needed to definitively establish whether or not the spin resonance is a singlet-triplet excitation (in some or all of the Fe-based superconductors).

The meaning of the resonance can be examined further by various scaling approaches.  One method of scaling is to simply scale the spin resonance energy by T$_C$.  The scaling of the spin resonance to k$_B$T$_C$ and, where available, to twice the maximum superconducting gap is shown in Table \ref{reson}.  The scaling with T$_C$ in the cuprates is typically quoted to be around $5k_BT_C$ \cite{Hufner2008}.  The scaling with T$_C$ observed in the Fe-based superconductors, $\sim$4.9 k$_B$T$_C$ (see Table \ref{reson}), is consistent with that in the cuprates.  In contrast in the heavy Fermion materials such scaling is not applicable as exemplified by CeCoIn$_5$ (values included in Table \ref{reson} for reference) \cite{Stock2008}.  Stock, \textit{et al.} \cite{Stock2008} point this out and suggested an alternative scaling whereby the resonance energy is scaled by twice the maximum superconducting gap (E$_r$/2$\Delta$) (note: A related scaling was also suggested by Mourachkine\cite{Mourachkine1999}).  This scaling gives agreement for three different d-wave superconductors, Bi$_2$Sr$_2$CaCu$_2$O$_{8+\delta}$, CeCoIn$_5$, and UPd$_2$Al$_3$ \cite{Stock2008}.  This scaling also was found to work in Ba$_{0.6}$K$_{0.4}$Fe$_2$As$_2$\cite{Christianson2008}.
Subsequently, the scaling of the resonance energy with the gap has been examined systematically in a large number of unconventional superconductors, including several Fe based superconductors, yielding a universal value of $\sim0.64$ \cite{Yu_Greven2009}.  As shown in Table \ref{reson}, in cases where the gap is available, this scaling is also observed in the Fe-based materials.  This provides a graphic demonstration of the link between the superconducting gap and spin fluctuations in unconventional superconducting materials.

A commonly held view of the superconducting resonance is that the enhancement of intensity in the INS spectra is due to a sign change of the superconducting order parameter on different parts of the Fermi surface.   Thus, in the cuprates and heavy Fermion materials the observation of a spin resonance is typically taken as evidence for d-wave superconductivity. Early theoretical investigations of Fe-based superconductors suggested an $s_\pm$ wave pairing symmetry \cite{Mazin2008a,Kuroki2008}.  The observation of a spin resonance in Ba$_{0.6}$K$_{0.4}$Fe$_2$As$_2$ \cite{Christianson2008} where photoemission measurements found no nodes in the superconducting gap \cite{Ding2008} are naturally explained within an $s_\pm$ picture. Moreover, the energy of the resonance seems to agree reasonably with theoretical calculations for the neutron spectra \cite{Maier2008,Korshunov2008a,Maier2009} in a number of materials \cite{Christianson2008,Lumsden2009,Inosov2010}.  An alternative point of view has been presented by Onari, \textit{et al.}\cite{Onari2010}  They argue that a sign change in the superconducting order parameter is not needed to give rise to a peak in the dynamic susceptibility and that the spin resonance feature can be understood assuming $s_{++}$ pairing symmetry.  Future experimental and theoretical investigation of this proposal would be interesting.

\section{Summary}

The first evidence of the interplay between magnetism and superconductivity in the Fe-based superconductors was the presence of magnetism in the concentration dependent phase diagrams.  The parent compounds exhibit a magnetically ordered state which is suppressed with doping and superconductivity appears at higher concentrations.  While this general behavior is common for different materials, the behavior near the boundary between superconductivity and magnetism is material specific. The magnetically ordered state of some materials vanish abruptly with the appearance of superconductivity, in others superconductivity emerges precisely as the magnetic order is destroyed, and still others exhibit coexistence between the magnetically ordered and superconducting states.  This issue of microscopic phase coexistence has been most carefully examined in doped BaFe$_2$As$_2$ with most measurements supporting phase coexistence in Ba(Fe$_{1-x}$Co$_x$)$_2$As$_2$ while phase separation seems to occur in Ba$_{1-x}$K$_x$Fe$_2$As$_2$.

The magnetic structure of the parent compounds indicate an identical in-plane spin arrangement in the 1111, 122, and 111 materials although the stacking of neighboring planes along the $c$-axis is material dependent.  This stripe-like structure consists of moments oriented along the orthorhombic $a$-axis stacked antiferromagnetically along $a$ and ferromagnetically along $b$.  Several explanations of this magnetic structure have been proposed including Fermi surface nesting, local moments interacting via frustrated superexchange interactions, local moments interacting via longer range interactions primarily of an itinerant nature, and orbital ordering.  The ordered Fe moments are very small in the RFeAsO parent compounds and, interestingly, $^{57}$Fe M\"{o}ssbauer measurements indicate a very similar magnetic moment for all measured rare earths while neutron diffraction indicates very different behavior particularly for the case of R=Ce where a much larger moment is observed (a similar effect is also seen in $\mu$SR).  Curiously, the exact opposite behavior is seen in the AFe$_2$As$_2$ parent compounds where neutron diffraction shows a very similar Fe moment for all values of A while $^{57}$Fe M\"{o}ssbauer shows strong variation with A with very different internal fields observed for A=Ba and A=Sr,Eu.  The magnetic structure of the 11 materials is a particularly interesting case as the magnetic structure is found to be quite different than the 1111 and 122 materials despite a very similar Fermi surface topology seemingly inconsistent with a picture of magnetic order occurring purely as a result of a nesting instability.

While many of the details still remain unknown the investigations of the spin excitations described above have already established much of the basic behavior.  Namely, that the interactions in the magnetically ordered parent compounds have been determined to be anisotropic 3D interactions.  With doping the correlations along the c-axis appear to be more quickly suppressed so that in optimally doped samples where no long range magnetic order is present the spin excitations appear to be 2D in character much as in the normal state of the parent compounds.  This latter behavior is not entirely unexpected from the magnetic phase diagrams of the Fe-based materials since the paramagnetic state of the parent compounds appears to be the same paramagnetic state out of which superconductivity develops in the optimally doped superconducting materials.   In the superconducting materials, spin fluctuations near (0.5 0.5 0)$_T$ have been found in all cases investigated.  These investigations further show the development of a spin resonance below T$_C$ demonstrating a direct interplay between magnetism and superconductivity.   For now the observation of the spin resonance in the Fe-based materials appears to support a $s_\pm$ pairing symmetry, but this issue is far from settled.   In the few cases studied by INS the spin fluctuations seem to weaken considerably and disappear in the overdoped regime.  Certainly much work remains to be done in these fascinating materials.

\section{Acknowledgements}

The authors would like to extend our sincere thanks to all of our collaborators.  In particular we thank E. Goremychkin, M. McGuire, T. Maier, D. Mandrus, S. Nagler, R. Osborn, B. Sales, A. Sefat, and D. Singh.  We would also like to thank our colleagues who have graciously allowed us to reproduce their work here.  Work at ORNL was supported by the Scientific User Facilities Division Office of Basic Energy Sciences, DOE.

\end{document}